\documentclass[aps,superscriptaddress,showpacs,nofootinbib,eqsecnum]{revtex4} 
  
\usepackage{epsfig}  
\input{epsf}  
  
\begin{document}  
  
\title{Effective Potential and Thermodynamics for a Coupled Two-Field Bose Gas
  Model}
  
\author{Marcus Benghi Pinto} \affiliation{Departamento de F\'{\i}sica,
  Universidade Federal de Santa Catarina, 88040-900 Florian\'{o}polis, Santa
  Catarina, Brazil}
  
\author{Rudnei O. Ramos} \affiliation{Departamento de F\'{\i}sica Te\'orica,
  Universidade do Estado do Rio de Janeiro, 20550-013 Rio de Janeiro, RJ,
  Brazil}
  
\author{Frederico F. de Souza Cruz} \affiliation{Departamento de F\'{\i}sica,
  Universidade Federal de Santa Catarina, 88040-900 Florian\'{o}polis, Santa
  Catarina, Brazil}
  
\begin{abstract}  
    
  We study the thermodynamics of a two-species homogeneous and dilute Bose gas
  that is self-interacting and quadratically coupled to each other.  We make
  use of field theoretical functional integral techniques and evaluate the
  one-loop finite temperature effective potential for this system considering
  the resummation of the leading order temperature dependent as well as
  infrared contributions. The symmetry breaking pattern associated to the
  model is then studied by considering different values of self and
  inter-species couplings. We pay special attention to the eventual appearance
  of re-entrant phases and/or shifts in the observed critical temperatures as
  compared to the mono atomic (one-field Bose) case.
  
\end{abstract}  
  
  
\maketitle
  
\section{Introduction}  
\label{sec1}  
  
Due to a larger region of parameters, multi-field theories may exhibit a much
richer phase structure in comparison to single field theories. Phenomena that
are completely absent in single field theories may arise in some regions of
the space parameter related to multi-field theories.  {}From a qualitative
point of view one expects that, at finite temperatures and/or densities,
multi-field theories will display phase transition patterns which cannot occur
in the single field case (see for instance Refs. \cite{nossoprd,prdjulia} and
references therein).  Quantitatively, on the other hand, the actual value of
critical quantities may be different in the two cases. {}For example, consider
a scalar multi-field theory composed by two different fields, $\Phi$ and
$\chi$, which self interact via quartic interactions (e.g., $g_\Phi \Phi^4$
and $g_\chi \chi^4$). The two fields may also interact quadratically with each
other, e.g., with an interaction term $g \Phi^2 \chi^2$. In general, these
type of interactions lead to an $O(N_\Phi) \times O(N_\chi)$ invariant theory,
where $N_\Phi$ and $N_\chi$ represent the number of components of each field.
In the relativistic case one may think of $\Phi$ and $\chi$ as representing an
extended Higgs sector of the standard model, e.g.  the Kibble-Higgs sector of
a $SU(5)$ grand-unified theory, with $N_\Phi=90$ and $N_\chi=24$
\cite{bimonte}. In the non-relativistic case one can associate $\Phi$ and
$\chi$ to two different species of bosonic atoms in an homogeneous dilute Bose
gas with $N_\Phi=N_\chi=2$. To assure stability the numerical values of the
couplings need to observe certain constraints whose general form is $ g_\Phi\,
g_\chi > b\, g^2$, where the self-couplings satisfy $g_\Phi > 0$, $g_\chi >
0$, whereas $g$ can be either positive or negative.  $b$ is some positive real
number whose value depends on the way the interactions are normalized in the
Lagrangian density. At the same time, stability (boundness of the potential
energy) only requires that the coupling constant be positive in the one field
case. It is then easy to understand what was said above about the numerical
values of critical quantities, like the critical temperature, $T_c$, which in
many cases depends on the masses and couplings, causing $T_c^{\rm multi} \ne
T_c^{\rm mono}$. Also, the way the transitions occur can be highly influenced
by the presence of the crossed interaction term proportional to $g$, as first
noted by Weinberg \cite{weinberg}. This fact arises as a consequence of $g$
assuming either positive or negative values and still keep the theory bounded
from below.  In his work, Weinberg used perturbation theory to analyze a
relativistic $O(N_\Phi)\times O(N_\chi)$ model at finite temperature. He found
that, for $g<0$, one may find regions of the parameter space where unexpected
phenomena arise. In particular, it was found that a symmetry that was broken
at $T=0$ could remain broken at arbitrarily high temperatures, in what was
called symmetry nonrestoration (SNR). Also, a symmetry that was unbroken at
$T=0$ could be broken at some finite $T_c$ in a manifestation of inverse
symmetry breaking (ISB). It is worth recalling that these two phenomena never
show up in the mono-field $O(N)$ theories where one always reaches the
symmetric phase at some finite temperature.
  
As phenomena happening at finite temperatures, one could argue that SNR and
ISB are just artifacts of perturbation theory, which although used in Ref.
\cite {weinberg}, is well known to be inadequate to treat high temperature
field theories \cite{kapusta,lebellac}.  However, powerful non-perturbative
techniques \cite{nossoprd,roos,eles} that include resummations of leading
order and infrared terms and thermal effects on the couplings, have confirmed
the possibility of these phenomena showing up in the relativistic case. The
nonrelativistic case, where the appearance of such phenomena is highly counter
intuitive, has been treated very recently by some of the present authors
\cite{prdjulia,jpa1,jpa2}. There, it was shown that SNR and ISB cannot
manifest themselves when thermal effects on the couplings are taken into
account. A broken symmetry at $T=0$ is always restored at some finite $T_c$
while an unbroken symmetry (at $T=0$) can be broken at a finite $T_c^{\rm
  ISB}$ only to be restored at a higher critical temperature characterizing a
re-entrant phase that is typical of many condensed matter systems, like the
Rochelle salt, spin glasses, compounds known as the manganites, liquid
crystals and many others materials, as recently reviewed in \cite{inverse}.
In the work carried out in Ref. \cite {prdjulia} one body terms proportional
to $\kappa_\Phi \Phi^2$ (and $\kappa_\chi \chi^2$) have been introduced to
drive symmetry breaking. Then, apart from the masses and couplings one has the
parameters $\kappa$ which may, for example, represent external fields that are
temperature independent at the tree level.
  
The aim of the present work is also to analyze a multi-field nonrelativistic
theory but this time we want to make contact with the case that is relevant
for dilute homogenous weakly interacting Bose gases. Therefore, we will choose
a $U(1) \times U(1)$ version (or equivalently, $O(2) \times O(2)$) taking the
one body parameters as representing the chemical potentials. As we shall see
this choice makes the finite temperature treatment very complex and here we
will use functional techniques to evaluate the effective potential (or free
energy). This will allow us to investigate the possibility of SNR/ISB
occurring in coupled homogeneous Bose gases. At the same time we will be in
position to check if the coupling of two different species of Bose gases
shifts the critical temperatures in relation to the case where this coupling
does not exist. Due to the non-perturbative nature of such evaluation we will
consider a resummation of the leading order temperature dependent and infrared
contributions to the effective potential to one-loop. As we shall see, the
present investigation excludes the possibility of re-entrant phases when the
nonrelativistic models studied in Ref. \cite {nossoprd} contain the chemical
potentials needed to represent Bose gases.  Moreover, the values of the
critical temperatures for the two different species seem to be insensitive to
the existence of the (new) cross coupling. As in the mono atomic case, their
values only depend on the density (and mass) of each specie coinciding with
the ideal gas situation. Although the conclusion about the non existence of
re-entrant phases is very plausible the statement about the critical
temperature should be taken with care. This is because here we are only
resumming one-loop (or direct contributions) which do not contribute at the
critical point due to the Hugenholtz-Pines theorem. On the other hand, our
non-perturbative calculation can reveal the differences in the thermal
behavior of uncoupled and coupled Bose gases at least in the region $0 \le T <
T_c$. The point $T=T_c$ can be fully exploited only by using non-perturbative
techniques which go beyond one-loop \cite{shift}.

The experimental realization of the Bose-Einstein condensation (BEC) in dilute
atomic gases has greatly stimulated an enormous number of theoretical studies
in this field (for recent reviews on the theory and experiments, see for
instance Ref. \cite{becreview}). Most of this interest comes from the fact
that in these experiments a great deal of control can be achieved in almost
every parameter of the system. Thus, experiments in dilute atomic Bose gases
provide a perfect ground to test numerous models and field theory methods
applied to these models, as for example finite temperature quantum field
theory methods as commonly used to study phase transition properties of
relativistic models. This makes BEC one of the most attractive and promising
systems in which one can use models and can test schemes and approximations
that could also prove useful in very different environments such as in the
early universe, heavy-ion collisions, etc, where the use of one or multi-field
models may have importance in their understanding.  Therefore, the extension
to BEC systems of the analysis performed in previous works in the phase
structure of multi-field models is particularly interesting given the
possibility of using experiments in dilute atomic Bose gases as an analog
system to model and test these finite temperature quantum field theory
systems.

This work is organized as follows. In Sec. II we present the model
representing two, self-interacting, coupled Bose gases. In Sec. III we
evaluate the finite temperature effective potential to one-loop order. Section
IV, which is divided in two subsections, presents the evaluation of
thermodynamic quantities as well as self-energies.  In the first subsection we
consider the one-field case whereas the two-field case is considered in the
next subsection. In Sec. V we consider the high temperature (symmetry
restored) phases. Our conclusions and final remarks are presented in Sec. VI.

\section{The Two self-interacting coupled Model for Bose gases}  
  
Let us now consider the case of two coupled Bose gases in the presence of
self-interactions.  The model we consider is similar to the ones used in other
theoretical studies of homogeneous dilute coupled Bose gases \cite{coupled},
that consists of a hard core sphere gas model described by nonrelativistic
interacting (complex) scalar fields, with an overall repulsive potential. This
system can be described by the following global $U_\Psi(1) \times U_\Phi(1)$
invariant Lagrangian model for two-species nonrelativistic complex scalar
field $\Phi$ and $\Psi$, with self-couplings $g_\Phi$ and $g_\Psi$ and
inter-species coupling $g$,
  
\begin{eqnarray}  
{\cal L}(\Phi^*,\Phi,\Psi^*,\Psi) &=&  
\Phi^* \left(i \partial_t + \frac{1}{2m_\Phi}\nabla^2  
\right) \Phi + \mu_\Phi \Phi^* \Phi   
- \frac{g_\Phi}{2} (\Phi^* \Phi)^2  
\nonumber \\  
&+& \Psi^* \left(i \partial_t + \frac{1}{2m_\Psi}\nabla^2  
\right) \Psi + \mu_\Psi \Psi^* \Psi   
- \frac{g_\Psi}{2} (\Psi^* \Psi)^2  
- g (\Phi^* \Phi) (\Psi^* \Psi)\;,  
\label{NRL}  
\end{eqnarray}

\noindent  
where the associated chemical potentials are represented by $\mu_i$ ($i=\Psi$
or $\Phi$) while $m_i$ represent the masses. {}For the hard core sphere
self-interactions we take the phenomenological coupling constants as being the
ones normally used in the absence of cross interactions and which are valid in
the dilute gas approximation \cite{becreview}. In terms of the corresponding
$s$-wave scattering lengths, $a_i$, they can be written as $g_i = 4 \pi
a_i/m_i$, while the cross-coupling is chosen as $g=4 \pi
a_{\Psi,\Phi}/m_{\Psi,\Phi}$ where $m_{\Psi,\Phi}=m_\Psi m_\Phi/(m_\Psi +
m_\Phi)$ represents a reduced mass.
  
The use of analog condensed matter systems to study multi-field theory models
like (\ref{NRL}), can be envisaged by the use of a system composed by a
mixture of coupled atomic gases, like the ones originally produced in Ref.
\cite{myatt} in which one has the same chemical element in two different
hyperfine states and that may be treated as ``effectively distinguishable", or
just consider the mixing of two different mono-atomic Bose gases.

The equivalent finite temperature Euclidean ($\tau = i t$) spacetime action to
Eq. (\ref{NRL}) is given by

\begin{eqnarray}  
\lefteqn{ \!\!\!\!\! \!\!\!\! S_E (\beta) \! = \!\! \int_0^\beta \! d \tau \!   
\int \! d^3 x \! \left[\!  
\Psi^* \! \left( \frac{\partial}{\partial \tau} \! - \!  
\frac{\nabla^2 }{2 m_\Psi} \!  - \! \mu_\Psi \right) \Psi \! + \!  
\frac {g_{\Psi}}{2}(\Psi^*\Psi)^2  
\right. }  \nonumber \\  
&&  \!\!\!\!\! \!\!\!\!\!\!\!\!\! + \left. \Phi^* \!\!   
\left( \! \frac{\partial}{\partial \tau} \!-  
\! \frac{\nabla^2 }{2 m_\Phi} \! - \! \mu_\Phi \! \right) \! \Phi \!   
+ \! \frac {g_{\Phi}}{2}(\Phi^*\Phi)^2 \!  
+ \! g  (\Psi^*\Psi)(\Phi^*\Phi) \right] ,  
\label{actionPsiPhi}  
\end{eqnarray}

\noindent  
where $T=1/\beta$ (we are considering throughout this paper all quantities in
natural unities, where $\hbar=k_{B} =1$).  Let us initially consider $\Psi$
and $\Phi$ as pure condensates, in the absence of fluctuations (pure
$c$-numbers).  By varying the action Eq. (\ref{actionPsiPhi}) with respect to
$\Psi$ and $\Phi$ we obtain the equations of motion (in Euclidean time),
analogous to the Gross-Pitaevskii equation in the case of a monoatomic gas,
given by
  
\begin{eqnarray}  
&& \partial_\tau \Psi = -\frac{1}{2 m_\Psi} \nabla^2 \Psi - \mu_\Psi \Psi  
+ \left( g_\Psi |\Psi|^2 + g |\Phi|^2 \right) \Psi \;, \nonumber \\  
&& \partial_\tau \Phi = -\frac{1}{2 m_\Phi} \nabla^2 \Phi - \mu_\Phi \Phi  
+ \left( g_\Phi |\Phi|^2 + g |\Psi|^2 \right) \Phi \;.  
\end{eqnarray}  
The minimization of the potential part of the action Eq. (\ref{actionPsiPhi})
leads to the relations involving the condensates for $\Psi$ and $\Phi$,
$\rho_{\Psi,c}= |\Psi|^2$ and $\rho_{\Phi,c}=|\Phi|^2$, respectively, and the
chemical potentials:
  
\begin{eqnarray}  
&&  \mu_\Psi =  g_\Psi \rho_{\Psi,c} + g \rho_{\Phi,c}  \;, \nonumber \\  
&&  \mu_\Phi = g_\Phi \rho_{\Phi,c} + g \rho_{\Psi,c}  \;.  
\label{mus}  
\end{eqnarray}  
In addition, for Eq. (\ref{mus}) to represent a local minimum of the
potential, we still need to make the requirement that it be bounded from
below, which requires the coupling constants to satisfy
  
\begin{equation}  
g_\Psi g_\Phi - g^2 >0\;,  
\label{stability}  
\end{equation}  
with repulsive self-interactions, $g_{\Psi} > 0$, $g_{\Phi} > 0$.  In the
experimental situation of a coupled binary system of Bose atoms, the condition
(\ref{stability}) is required for the mixture of condensates to be stable. In
the case where Eq. (\ref{stability}) is violated, de-mixing of the condensates
happens, so that the mixture of condensates will tend to separate spatially,
as actually observed in the experiment of the second reference in Ref.
\cite{myatt}.
  
Equation (\ref{mus}) also gives the condensate densities in terms of the
chemical potentials:
  
\begin{equation}  
\rho_{\Psi,c}= \frac{g_\Phi \mu_\Psi - g \mu_\Phi}{g_\Psi g_\Phi - g^2} \;,  
\label{rhoPsi2}  
\end{equation}  
and
  
\begin{equation}  
\rho_{\Phi,c}= \frac{g_\Psi \mu_\Phi - g \mu_\Psi}{g_\Psi g_\Phi - g^2} \;.  
\label{rhoPhi2}  
\end{equation}

\section{The One-loop Finite Temperature Effective Potential}

Let us now consider the effect of fluctuations. Here we choose to study the
system at equilibrium and will include fluctuations in the system by means of
the field theoretical method of the effective potential. We will follow
closely the functional integration derivation used by the authors in Ref.
\cite{norway} to the usual one field self-interacting Bose gas (for other
field theoretic methods applied to BEC problems, see also Ref. \cite{toms,gil}
and for a review Ref. \cite{jeans}).  As usual in the computation of the
effective potential, we start by decomposing the fields $\Phi$ and $\Psi$ in
Eq. (\ref{NRL}) in terms of (constant) background fields (which, without loss
of generality, can be taken as real fields, $c$-numbers) $\phi_0$ and
$\psi_0$, respectively, and ($q$-number) fluctuations $\Phi$ and $\Psi$, which
in terms of real components, become

\begin{eqnarray}  
&&\Phi = \frac{1}{\sqrt{2}}\left( \phi_0 + \phi_1 + i \phi_2 \right)  
\;,\label{Phi} \\  
&& \Psi = \frac{1}{\sqrt{2}}\left( \psi_0 + \psi_1 + i \psi_2 \right)  
\;.\label{Psi}  
\end{eqnarray}  
  
\noindent  
Considering the leading order in the fluctuations (which is equivalent to keep
terms up to order $\hbar$ in the effective potential, or one-loop order), when
substituting Eqs. (\ref{Phi}) and (\ref{Psi}) in Eq. (\ref{NRL}) we only need
to keep the quadratic terms in the fluctuation fields for the computation of
the one-loop potential for the background fields, $\phi_0$ and $\psi_0$. We
then obtain the functional partition function to one-loop order
  
\begin{equation}  
Z[\beta] = \int D\psi_1 D \psi_2 D \phi_1 D\phi_2  
\exp\left( - S_2   
\right) \;,  
\label{functional}  
\end{equation}  
  
\noindent  
where the functional integral is restricted over fields satisfying the
periodic boundary conditions (the Kubo-Martin-Schwinger condition) $\Psi ({\bf
  x},\tau) = \Psi({\bf x},\tau+\beta)$ and $\Phi ({\bf x},\tau) = \Phi({\bf
  x},\tau+\beta)$. $S_2$ is the Euclidean action to quadratic order in the
fluctuation fields,

\begin{equation}  
S_2 = \int_0^\beta d \tau \int d^3 x \left[ -\frac{\mu_\Phi}{2} \phi_0^2 +  
\frac{g_\Phi}{8} \phi_0^4 - \frac{\mu_\Psi}{2} \psi_0^2 +  
\frac{g_\Psi}{8} \psi_0^4 + \frac{g}{4} \phi_0^2 \psi_0^2  
+ \frac{1}{2} \chi \cdot \hat{M} \cdot \chi \right]\;,  
\label{LE}  
\end{equation}  
  
\noindent  
where we have defined the vector $\chi=(\phi_1,\phi_2,\psi_1,\psi_2)$ and
$\hat{M}$ is the matrix operator for the quadratic terms in the fluctuations,
  
\begin{equation}  
\hat{M} =   
\left(  
\begin{array}{cccc}  
\frac{-\nabla^2}{2 m_\Phi} - \mu_\Phi +\frac{3g_\Phi}{2} \phi_0^2 +  
\frac{g}{2} \psi_0^2 & i \partial_\tau & g \phi_0 \psi_0 & 0 \\  
- i \partial_\tau & \frac{-\nabla^2}{2 m_\Phi} - \mu_\Phi +  
\frac{g_\Phi}{2} \phi_0^2 + \frac{g}{2} \psi_0^2 & 0 & 0 \\  
g \phi_0 \psi_0 & 0 & \frac{-\nabla^2}{2 m_\Psi} - \mu_\Psi +  
\frac{3g_\Psi}{2}  
\psi_0^2 + \frac{g}{2} \phi_0^2 & i \partial_\tau \\  
0 & 0 &  - i \partial_\tau & \frac{-\nabla^2}{2 m_\Psi} - \mu_\Psi +  
\frac{g_\Psi}{2} \psi_0^2 + \frac{g}{2} \phi_0^2  
\end{array}  
\right).  
\label{Mmatrix}  
\end{equation}  
  
\noindent  
The partial time derivative in (\ref{Mmatrix}) is over Euclidean time:
$\partial_\tau = \partial/\partial\tau$, $\tau = i t$.  As usual, the
effective potential is defined from the functional partition function by
  
\begin{equation}  
V_{\rm eff} (\psi_0,\phi_0) = -\frac{1}{\beta V} \ln Z[\beta]\;,  
\end{equation}  
where $V$ is the volume.  By performing the functional integration in the
quadratic fluctuations $\chi$, the one-loop effective potential $V_{\rm
  eff}(\phi_0,\psi_0)$ obtained from Eq. (\ref{LE}) is then given by
  
\begin{eqnarray}  
V_{\rm eff}(\phi_0,\psi_0) =   
-\frac{\mu_\Phi}{2} \phi_0^2 +  
\frac{g_\Phi}{8} \phi_0^4 - \frac{\mu_\Psi}{2} \psi_0^2 +  
\frac{g_\Psi}{8} \psi_0^4 + \frac{g}{4} \phi_0^2 \psi_0^2  
+ \frac{1}{2} \ln \det \hat{M} \;,  
\label{Veff}  
\end{eqnarray}  
  
\noindent  
where the last term on the RHS of Eq. (\ref{Veff}) comes from the functional
integral over the components of $\chi$,
  
\begin{equation}  
\frac{1}{2} \ln \det \hat{M} = -\frac{1}{\beta V} \int  
D \phi_1 D \phi_2 D \psi_1 D \psi_2 \exp\left[  
- \int_0^\beta d \tau \int d^3 x \left(  
\frac{1}{2} \chi \cdot \hat{M} \cdot \chi \right) \right]\;.  
\label{funcint}  
\end{equation}  
  
\noindent  
Expressing Eqs. (\ref{funcint}) and (\ref{Mmatrix}) in the space-time momentum
{}Fourier transform form, we obtain
   
\begin{eqnarray}  
\frac{1}{2} \ln \det \hat{M} &=& \frac{1}{2} \frac{1}{\beta}  
\sum_{n=-\infty}^{+\infty} \int \frac {d^3 {\bf q}}{(2 \pi)^3} \ln \left\{  
\left[ \omega_n^2 + E_\Phi^2({\bf q}) \right]   
\left[ \omega_n^2 + E_\Psi^2({\bf q}) \right]   
\right. \nonumber \\  
&-& \left.  g^2 \phi_0^2 \psi_0^2 \left[ \omega_\Phi({\bf q}) +   
\frac{g_\Phi}{2} \phi_0^2 + \frac{g}{2} \psi_0^2 \right]  
\left[ \omega_\Psi({\bf q}) +   
\frac{g_\Psi}{2} \psi_0^2 + \frac{g}{2} \phi_0^2 \right]  
\right\}\;,  
\label{lndet}  
\end{eqnarray}  
  
\noindent  
with
  
\begin{equation}  
E_\Psi({\bf q}) = \sqrt{\left[ \omega_\Psi({\bf q}) +   
\frac{3g_\Psi}{2} \psi_0^2 + \frac{g}{2} \phi_0^2 \right]  
\left[ \omega_\Psi({\bf q}) +   
\frac{g_\Psi}{2} \psi_0^2 + \frac{g}{2} \phi_0^2 \right]}\;,  
\label{EPsi}  
\end{equation}and  
  
\begin{equation}  
E_\Phi({\bf q}) = \sqrt{\left[ \omega_\Phi({\bf q}) +   
\frac{3g_\Phi}{2} \phi_0^2 + \frac{g}{2} \psi_0^2 \right]  
\left[ \omega_\Phi({\bf q}) +   
\frac{g_\Phi}{2} \phi_0^2 + \frac{g}{2} \psi_0^2 \right]}\;,  
\label{EPhi}  
\end{equation}  
  
\noindent  
where $\omega_i({\bf q})$, $i=\Phi,\Psi$ is given by
  
\begin{equation}  
\omega_i({\bf q})= \frac{{\bf q}^2}{2m_i} - \mu_i\;,  
\label{omegaq}  
\end{equation}  
  
\noindent  
and $\omega_n$ in Eq. (\ref{lndet}) represents the Matsubara frequencies,
$\omega_n = 2 \pi n/\beta$, $n=0,\pm 1, \pm 2, \cdots$.
  
Considering the classical condensates densities in the absence of
fluctuations, Eqs. (\ref{rhoPsi2}) and (\ref{rhoPhi2}), noticing that
$\rho_{\Psi,c} = \psi_0^2/2$ and $\rho_{\Phi,c} = \phi_0^2/2$ and substituting
them in Eqs. (\ref{EPsi}) and (\ref{EPhi}) we recover the Bogoliubov
dispersion relations for the gases in the broken phase

\begin{eqnarray}  
E_\Psi &=& \sqrt{\left[ \omega_\Psi({\bf q}) +   
3g_\Psi \rho_{\Psi,c} + g \rho_{\Phi,c} \right]  
\left[ \omega_\Psi({\bf q}) +   
g_\Psi \rho_{\Psi,c} + g \rho_{\Phi,c} \right]} \nonumber \\  
&=& \sqrt{ \frac{{\bf q}^2}{2m_\Psi} \left( \frac{{\bf q}^2}{2m_\Psi} +   
2 g_\Psi \rho_{\Psi,c} \right) } \;,  
\label{EPsi0}  
\end{eqnarray}  
and
  
\begin{eqnarray}  
E_\Phi &=& \sqrt{\left[ \omega_\Phi({\bf q}) +   
3g_\Phi \rho_{\Phi,c} + g \rho_{\Psi,c} \right]  
\left[ \omega_\Phi({\bf q}) +   
g_\Phi \rho_{\Phi,c} + g \rho_{\Psi,c} \right]} \nonumber \\  
&=& \sqrt{ \frac{{\bf q}^2}{2m_\Phi} \left( \frac{{\bf q}^2}{2m_\Phi} +   
2 g_\Phi \rho_{\Phi,c} \right) } \;,  
\label{EPhi0}  
\end{eqnarray}  
  
\noindent  
which are consequences of the breaking of the two continuous symmetries of the
model and Goldstone's theorem. We can then recognize in Eqs. (\ref{EPsi0}) and
(\ref{EPhi0}) a Higgs and Goldstone modes like terms, that we denote
respectively by $H_i$ and $G_i$, given by
  
\begin{eqnarray}  
H_\Psi({\bf q},\phi_0,\psi_0) &=&  \omega_\Psi({\bf q}) + \frac{3g_\Psi}{2} \psi_0^2 +   
\frac{g}{2} \phi_0^2\;,   
\label{HPsi}\\  
G_\Psi({\bf q},\phi_0,\psi_0) &=&  \omega_\Psi({\bf q}) + \frac{g_\Psi}{2} \psi_0^2 +   
\frac{g}{2} \phi_0^2\;,   
\label{GPsi}\\  
H_\Phi({\bf q},\phi_0,\psi_0) &=&  \omega_\Phi({\bf q}) + \frac{3g_\Phi}{2} \phi_0^2 +   
\frac{g}{2} \psi_0^2\;,   
\label{HPhi}\\  
G_\Phi({\bf q},\phi_0,\psi_0) &=&  \omega_\Phi({\bf q}) + \frac{g_\Phi}{2} \phi_0^2 +   
\frac{g}{2} \psi_0^2  
\label{GPhi}\;.  
\end{eqnarray}  
  
In terms of Eqs. (\ref{HPsi})-(\ref{GPhi}), the contribution of fluctuations
to the classical potential in (\ref{Veff}), $\Delta V$, becomes
  
\begin{eqnarray}  
\Delta V &=&  \frac{1}{2} \ln \det \hat{M} = \frac{1}{2} \frac{1}{\beta}  
\sum_n \int \frac {d^3 {\bf q}}{(2 \pi)^3} \ln \left[  
\left( \omega_n^2 + H_\Psi G_\Psi \right) \left( \omega_n^2 + H_\Phi G_\Phi \right)  
- g^2 \psi_0^2 \phi_0^2 G_\Psi G_\Phi \right] \nonumber \\  
&=& \frac{1}{2} \frac{1}{\beta}  
\sum_n \int \frac {d^3 {\bf q}}{(2 \pi)^3} \ln \left[  
\left( \omega_n^2 + A^2 \right) \left( \omega_n^2 + B^2 \right) \right]   
\nonumber \\  
&=& \frac{1}{2} \frac{1}{\beta}  
\sum_n \int \frac {d^3 {\bf q}}{(2 \pi)^3} \ln \left( \omega_n^2 + A^2 \right)  
+ \frac{1}{2} \frac{1}{\beta}  
\sum_n \int \frac {d^3 {\bf q}}{(2 \pi)^3} \ln \left( \omega_n^2 + B^2 \right)\;,  
\label{DeltaV}  
\end{eqnarray}  
  
\noindent  
where the terms $A$ and $B$ in Eq. (\ref{DeltaV}) are given by
  
\begin{equation}  
A^2 = \frac{H_\Psi G_\Psi + H_\Phi G_\Phi}{2} -  
\frac{1}{2} \left[ \left( H_\Psi G_\Psi - H_\Phi G_\Phi \right)^2 +  
4 g^2 \Psi^2 \Phi^2 G_\Psi G_\Phi \right]^{1/2}\;,  
\label{A}  
\end{equation}  
and
  
\begin{equation}  
B^2 = \frac{H_\Psi G_\Psi + H_\Phi G_\Phi}{2} +  
\frac{1}{2} \left[ \left( H_\Psi G_\Psi - H_\Phi G_\Phi \right)^2 +  
4 g^2 \Psi^2 \Phi^2 G_\Psi G_\Phi \right]^{1/2}\;.  
\label{B}  
\end{equation}  
  
\noindent  
The one-loop correction to the classical potential when expressed in the form
of the last term in the RHS of Eq. (\ref{DeltaV}) is a suitable form that
allows to easily perform the sum over the Matsubara frequencies by using the
formula,
  
\begin{equation}  
\frac{1}{\beta} \sum_{n=-\infty}^{+\infty} \frac{1}{\omega_n^2 + \omega^2}  
= \frac{1}{2 \omega} \left( 1 + \frac{2}{e^{\beta \omega}-1}\right) \;.  
\label{identity}  
\end{equation}  
  
\noindent  
Defining the quantity $v(\omega)$,
  
\begin{equation}  
v(\omega) = \frac{1}{\beta}  \sum_{n=-\infty}^{+\infty}\ln (\omega_n^2   
+\omega^2) \;,  
\end{equation}  
  
\noindent  
we have
  
\begin{equation}  
\frac{\partial v(\omega)}{\partial \omega} = \frac{1}{\beta}  
\sum_{n=-\infty}^{+\infty} \frac{2 \omega}{\omega_n^2 + \omega^2} \,\,,  
\end{equation}  
  
\noindent  
which, from Eq. (\ref{identity}), becomes
  
\begin{equation}  
\frac{\partial v(\omega)}{\partial \omega} =1 + 
\frac{2}{e^{\beta \omega}-1}\,\,,  
\end{equation}  
  
\noindent  
and so
  
\begin{equation}  
v(\omega) = \omega + \frac{2}{\beta} \ln \left(1-e^{-\beta \omega}\right)  
+\;{\rm terms\; independent\; of \;}\omega \;.  
\end{equation}  
  
\noindent  
Neglecting the constant terms, we then obtain for Eq. (\ref{DeltaV}) the
result
  
\begin{equation}  
\Delta V = \frac{1}{2} \int \frac {d^3 {\bf q}}{(2 \pi)^3}  
\left(A + B \right) +  
\frac{1}{\beta} \int \frac {d^3 {\bf q}}{(2 \pi)^3} \left[  
\ln \left( 1- e^{- \beta A} \right) + \ln \left( 1- e^{- \beta B} \right)  
\right] \;.  
\label{DeltaV2}  
\end{equation}  
  
\section{Thermodynamic Quantities and Self-Energies}  
  
Given the effective potential, from Eqs. ({\ref{Veff}) and (\ref{DeltaV2}),

\begin{eqnarray}  
V_{\rm eff}(T,\phi_0,\psi_0) &=&   
-\frac{\mu_\Phi}{2} \phi_0^2 +  
\frac{g_\Phi}{8} \phi_0^4 - \frac{\mu_\Psi}{2} \psi_0^2 +  
\frac{g_\Psi}{8} \psi_0^4 + \frac{g}{4} \phi_0^2 \psi_0^2  
\nonumber \\  
&+& \frac{1}{2} \int \frac {d^3 {\bf q}}{(2 \pi)^3}  
\left(A + B \right)  
+ \frac{1}{\beta} \int \frac {d^3 {\bf q}}{(2 \pi)^3} \left[  
\ln \left( 1- e^{- \beta A} \right) + \ln \left( 1- e^{- \beta B} \right)  
\right]  \;,  
\label{Veff2}  
\end{eqnarray}  
  
\noindent  
with the functions $A\equiv A({\bf q},\phi_0,\psi_0) $ and $B\equiv B({\bf
  q},\phi_0,\psi_0)$, given by Eqs. (\ref{A}) and (\ref{B}), respectively, we
can compute all relevant thermodynamical functions. In particular, we have
that the pressure is defined as the negative of the effective potential
computed at its minima (which is the thermodynamical free energy of the
system),
  
\begin{equation}  
P\equiv P(T,\mu_\Phi,\mu_\Psi) = - V_{\rm eff} (T,\phi_0,\psi_0)   
\Bigr|_{\phi_0 = \phi_m, \psi_0=\psi_m}\;,  
\label{pressure}  
\end{equation}  
  
\noindent  
where $\phi_m$ and $\psi_m$ are the values of $\phi_0$ and $\psi_0$ that
extremizes (corresponding to a minimum of) the effective potential,
  
\begin{eqnarray}  
\frac{\partial V_{\rm eff} (T,\phi_0,\psi_0)}{\partial \phi_0}  
\Bigr|_{\phi_0 = \phi_m, \psi_0=\psi_m} =0\;,\;\;\;\;  
\frac{\partial V_{\rm eff} (T,\phi_0,\psi_0)}{\partial \psi_0}  
\Bigr|_{\phi_0 = \phi_m, \psi_0=\psi_m} =0\;.  
\label{minimum}  
\end{eqnarray}  
  
\noindent  
{}For the tree-level potential, $\psi_m$ and $\phi_m$ are given by Eqs.
(\ref{rhoPsi2}) and (\ref{rhoPhi2}), respectively (with $\rho_{\Psi,c} =
\psi_m^2/2$ and $\rho_{\Phi,c} = \phi_m^2/2$). As we will see below,
interactions will change these tree-level expressions and, consequently,
modifications will have to be implemented in the effective potential so as to
preserve the Goldstone's theorem in the presence of interactions and finite
temperature effects. {}From the pressure the total number of particles follows
as
  
\begin{eqnarray}  
\rho_\Psi = \frac{\partial P(T,\mu_\Phi,\mu_\Psi)}{\partial \mu_\Psi}  
\;,\;\;\;\;  
\rho_\Phi = \frac{\partial P(T,\mu_\Phi,\mu_\Psi)}{\partial \mu_\Phi}  
\;.  
\label{densities}  
\end{eqnarray}  
  
The corrections to the tree-level (zero temperature) densities change Eqs.
(\ref{rhoPsi2}) and (\ref{rhoPhi2}) according to Eq. (\ref{minimum}) and the
obtained expression for the effective potential. Let us see this with some
more details starting with the uncoupled gas case.

\subsection{The One-Field Case}

{}For $g=0$, Eqs. (\ref{A}) and (\ref{B}) become
  
\begin{equation}  
A^2 (g=0) =  \left( \omega_\Phi({\bf q}) + \frac{3g_\Phi}{2} \phi_0^2 \right)  
\left( \omega_\Phi({\bf q}) + \frac{g_\Phi}{2} \phi_0^2 \right)\;,  
\end{equation}  
and
  
\begin{equation}  
B^2 (g=0) =  \left( \omega_\Psi({\bf q}) + \frac{3g_\Psi}{2} \psi_0^2 \right)  
\left( \omega_\Psi({\bf q}) + \frac{g_\Psi}{2} \psi_0^2 \right)\;,  
\end{equation}  
  
\noindent  
and the contributions from the $\Phi$ and $\Psi$ fields to the effective
potential, Eq. (\ref{Veff2}), decouple. Lets consider then, e.g., the $\Phi$
field contribution. In this case

\begin{eqnarray}  
V_{\rm eff,\Phi}(T,\phi_0) &=&   
-\frac{\mu_\Phi}{2} \phi_0^2 +  
\frac{g_\Phi}{8} \phi_0^4   
\nonumber \\  
&+& \frac{1}{2} \int \frac {d^3 {\bf q}}{(2 \pi)^3}  
\sqrt{ \left( \omega_\Phi({\bf q}) + \frac{3g_\Phi}{2} \phi_0^2 \right)  
\left( \omega_\Phi({\bf q}) + \frac{g_\Phi}{2} \phi_0^2 \right) }  
\nonumber \\  
&+& \frac{1}{\beta} \int \frac {d^3 {\bf q}}{(2 \pi)^3}   
\ln \left\{ 1- \exp\left[- \beta \sqrt{ \left( \omega_\Phi({\bf q}) +   
\frac{3g_\Phi}{2} \phi_0^2 \right)  
\left( \omega_\Phi({\bf q}) + \frac{g_\Phi}{2} \phi_0^2 \right)} \right] \right\} \;.  
\label{VeffPhi}  
\end{eqnarray}  
  
The zero temperature contribution in Eq. (\ref{VeffPhi}) is divergent and
require proper renormalization. It is easier to do it by performing the
momentum integral in $d=3-\epsilon$ dimensions and the resulting integral is
found to be finite in dimensional regularization (when taking $\epsilon \to 0$
at the end). This is so since the divergence in Eq. (\ref{VeffPhi}) is a
power-law one and, therefore, in dimensional regularization, the regularized
integral results to be finite.  We then obtain the minimum of the effective
potential as
  
\begin{eqnarray}  
\frac{\partial V_{\rm eff,\Phi} (T,\phi_0)}{\partial \phi_0}  
\Bigr|_{\phi_0 = \phi_m} =0 \Rightarrow  
\frac{\phi_m^2}{2} = \frac{\mu_\Phi}{g_\Phi} -\frac{1}{2}  
\int \frac {d^3 {\bf q}}{(2 \pi)^3}  \frac{ 2 \omega_\Phi({\bf q})   
+ \frac{3g_\Phi}{2} \phi_m^2 }  
{ \sqrt{ \left(  \omega_\Phi +   
\frac{3g_\Phi}{2} \phi_m^2 \right)  
\left(\omega_\Phi + \frac{g_\Phi}{2} \phi_m^2 \right)} }  
\left[ 1 + 2 n_\Phi ({\bf q}) \right]\;,  
\label{minimum2}  
\end{eqnarray}  
  
\noindent  
where
  
\begin{equation}  
n_\Phi ({\bf q}) =  \frac{1}{\exp\left[\beta   
\sqrt{ \left( \omega_\Phi({\bf q}) +   
\frac{3g_\Phi}{2} \phi_m^2 \right)  
\left( \omega_\Phi({\bf q}) +   
\frac{g_\Phi}{2} \phi_m^2 \right)} \right] -1 }\;,  
\label{BE1}  
\end{equation}  
  
\noindent  
is the Bose-Einstein distribution for the single self-interacting field and,
in Eq. (\ref{minimum2}), $\mu_\Phi/g_\Phi$ is the tree-level condensate
density (which also follows from Eq. (\ref{rhoPhi2}) for $g=0$).  By demanding
that the spectrum for the single self-interacting gas at finite temperature is
still gapless (Goldstone's theorem), we can define the condensate density like
for instance $\phi_m^2/2 = \bar{\mu}_\Phi/g_\Phi$, where $\bar{\mu}_\Phi$
denotes an effective chemical potential (we here will be using an analogous
definition as taken by the authors of Ref. \cite{norway} in their study of the
effective potential for a single self-interacting Bose field). This is
expected, since including the finite temperature contributions implies that
the original chemical potential must be changed accordingly\footnote{ The same
  effect happens in thermal field theory, where instead of a thermodynamic
  chemical potential, we would then now talk about a constant mass term for
  the field.  However, finite temperature contributions entering via the self-
  energies change this mass such as to make it temperature dependent.  This is
  equivalent to change the original mass by a ``dressed" one, where
  self-energy corrections are taken into account in the definitions of the
  field propagators.}.  In terms of these new definitions Eq. (\ref{minimum2})
becomes

\begin{eqnarray}  
\frac{\phi_m^2}{2} = \frac{\mu_\Phi}{g_\Phi} -  
\frac{1}{2} \int \frac {d^3 {\bf q}}{(2 \pi)^3}    
\frac{ 2\frac{{\bf q}^2}{2 m_\Phi}+  
\bar{\mu}_\Phi}  
{ \sqrt{ \frac{{\bf q}^2}{2 m_\Phi}  
\left(\frac{{\bf q}^2}{2 m_\Phi} +2\bar{\mu}_\Phi\right)} }  
\left\{ 1+ \frac{2}{\exp\left[\beta \sqrt{ \frac{{\bf q}^2}{2 m_\Phi}  
\left(\frac{{\bf q}^2}{2 m_\Phi} +  
2\bar{\mu}_\Phi\right)} \right] -1 } \right\}\;.  
\label{minimum3}  
\end{eqnarray}  
  
Replacing $\mu_\Phi$ by the effective chemical potential in the fluctuation
terms of the effective potential corresponds to replacing the tree-level field
propagators by the self-energy dressed ones. In this case, the equivalent of
the matrix $\hat{M}$, Eq. (\ref{Mmatrix}), for the one-field case, in the
basis $(\phi_1,\phi_2)$ and in momentum space, becomes
  
\begin{eqnarray}  
\hat{M}_\Phi(\omega_n,{\bf q}) =&& \left(  
\begin{array}{cc}  
\frac{{\bf q}^2}{2 m_\Phi} - \mu_\Phi +   
\frac{3g_\Phi}{2} \phi_0^2 + \Sigma_{\phi_1,\phi_1} &  -\omega_n   
+ \Sigma_{\phi_1,\phi_2}  \\  
\omega_n + \Sigma_{\phi_2,\phi_1}   & \frac{{\bf q}^2}{2 m_\Phi} - \mu_\Phi +   
\frac{g_\Phi}{2} \phi_0^2 + \Sigma_{\phi_2,\phi_2}  
\end{array}  
\right)\;.  
\label{MPhi}  
\end{eqnarray}  
One-loop diagrams that contribute to the self-energies are shown in {}Fig.
\ref{selfdiag}.
 
\begin{figure}[htb]  
  \vspace{0.5cm}
  \epsfig{figure=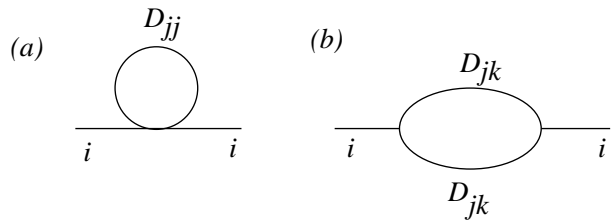,angle=0,width=8cm}
\caption[]{\label{selfdiag} Typical one-loop {}Feynman diagrams 
  contributing to the field self-energy. External lines stand for $\phi_i$ and
  the internal propagators $D_{jk}$ are defined by the inverse of the matrix
  of quadratic fluctuations $\hat{M}$. }
\end{figure}

Since $\Sigma_{\Phi_i,\Phi_j}$ for $i\neq j$ are linear in $\omega_n$ (see,
e.g. Ref. \cite{norway}), they are identically zero, $\Sigma_{\phi_1,\phi_2}
=\Sigma_{\phi_2,\phi_1}=0$.  {}From Eq. (\ref{MPhi}), the equivalent of the
Higgs and Goldstone mode terms, Eqs. (\ref{HPhi}) and (\ref{GPhi}) becomes
  
\begin{eqnarray}  
H_\Phi &=&   \frac{{\bf q}^2}{2 m_\Phi} - \mu_\Phi +   
\frac{3g_\Phi}{2} \phi_0^2 + \Sigma_{\phi_1,\phi_1}\;,   
\label{HPhi2}\\  
G_\Phi &=&  \frac{{\bf q}^2}{2 m_\Phi} - \mu_\Phi +   
\frac{g_\Phi}{2} \phi_0^2 + \Sigma_{\phi_2,\phi_2} \;.  
\label{GPhi2}  
\end{eqnarray}  
  
\noindent  
Therefore, Goldstone's theorem is preserved if we define the effective
potential $\bar{\mu}_\Phi$ as
  
\begin{equation}  
\bar{\mu}_\Phi = \mu_\Phi -  \Sigma_{\phi_2,\phi_2}\;,  
\label{barmu}  
\end{equation}  
  
\noindent  
and the minimum of the effective potential, $\phi_m$, is related with
$\bar{\mu}_\Phi$ by $\phi_m^2/2 = \bar{\mu}_\Phi/g_\Phi$.  Note that at the
critical temperature of Bose-Einstein condensation, $\bar{\mu}_\Phi=0$
($\phi_m=0$) and Eq. (\ref{barmu}) becomes $\mu_\Phi =
\Sigma_{\phi_2,\phi_2}$, which is just the standard form for the
Hugenholtz-Pines theorem \cite{jeans}.
  
The introduction of the self-energies $\Sigma_{\Phi_i,\Phi_i}$ can be
implemented at the Lagrangian density level right at the beginning in a
self-consistent way that avoids any possible overcounting of diagrams as would
be caused by the naive change of the matrix of quadratic fluctuations given by
Eq. (\ref{MPhi}). It can be checked that, without any further modifications,
the change produced by Eq. (\ref{MPhi}) starts overcounting diagrams at the
two-loop level (like two-bubble vacuum and self-energy diagrams in
perturbation theory).  This overcounting can be eliminated completely by
writing the original Lagrangian density, in terms of the field components
$\phi_1$ and $\phi_2$ like:
  
\begin{equation}  
{\cal  L} \to {\cal L} \left[\mu_\Phi\Phi_i^2 \to \bar{\mu}_\Phi\Phi_i^2 =   
(\mu_\Phi - \Sigma_{\Phi_i,\Phi_i})\Phi_i^2\right]  +  \frac{1}{2}  
\sum_i \Sigma_{\Phi_i,\Phi_i}\Phi_i^2\;,  
\label{newL}  
\end{equation}  
  
\noindent  
where the self-energy terms are at the same time added and subtracted in the
original action. While the added self-energies dress the field propagators,
the subtracted terms are treated as additional interaction terms and they here
act in the sense of subtracting the extra contributions coming from the
dressing of the Higgs and Goldstone modes, Eqs.  (\ref{HPhi2}) and
(\ref{GPhi2}). The procedure in (\ref{newL}) is common in various other
instances of studies involving resummation of quantum and temperature 
correction terms in quantum field theory \cite{sumsigma}.
  
{}From the self-energies, one can check that for temperatures $T \gg
\bar{\mu}_\Phi$, which is the regime of temperatures we are interested in, the
exchange diagrams (like the two-vertex one-loop diagrams of the form of {}Fig.
1b) are subleading compared to the tadpole diagrams, {}Fig. 1a. (see also Ref.
\cite{norway}).  In this case, $\Sigma_{\phi_1,\phi_1} \simeq
\Sigma_{\phi_2,\phi_2}$ and they are given, at the minimum of the effective
potential, by
  
\begin{equation}  
\Sigma_{\phi_1,\phi_1}(T) \simeq \Sigma_{\phi_2,\phi_2}(T)   
\simeq  
\frac{g_\Phi}{2} \int \frac {d^3 {\bf q}}{(2 \pi)^3}    
\frac{ 2\frac{{\bf q}^2}{2 m_\Phi}+  
\bar{\mu}_\Phi}  
{ \sqrt{ \frac{{\bf q}^2}{2 m_\Phi}  
\left(\frac{{\bf q}^2}{2 m_\Phi} +2\bar{\mu}_\Phi\right)} }  
\left\{ 1+ \frac{2}{\exp\left[\beta \sqrt{ \frac{{\bf q}^2}{2 m_\Phi}  
\left(\frac{{\bf q}^2}{2 m_\Phi} +  
2\bar{\mu}_\Phi\right)} \right] -1 } \right\}\;.  
\label{Sigma}  
\end{equation}  
   
\noindent  
Comparing Eq. (\ref{Sigma}) with (\ref{minimum3}), we get the result given
previously for the condensate density at high temperatures, $\phi_m^2/2=
\bar{\mu}_\Phi/g_\Phi$.
  
In terms of Eq. (\ref{MPhi}), the dressed effective potential therefore
becomes just like in Eq. (\ref{VeffPhi}), but with $\mu_\Phi$ exchanged by
$\bar{\mu}_\Phi$ in the correction terms for the tree-level potential. With
this self-energy ``improved" effective potential computed at the minimum
$\phi_0= \phi_m$, we obtain the pressure,
  
\begin{eqnarray}  
P(T,\mu_\Phi) &=& \frac{\mu_\Phi^2 - \Sigma_{\phi_2,\phi_2}^2(T)}{2 g_\Phi}  
-\frac{1}{2}  \int \frac {d^3 {\bf q}}{(2 \pi)^3}   
\sqrt{ \frac{{\bf q}^2}{2 m_\Phi}  
\left(\frac{{\bf q}^2}{2 m_\Phi} +  
2\bar{\mu}_\Phi\right)}   
\nonumber \\  
&-& \frac{1}{\beta} \int \frac {d^3 {\bf q}}{(2 \pi)^3}   
\ln \left\{ 1- \exp \left[ -\beta \sqrt{ \frac{{\bf q}^2}{2 m_\Phi}  
\left(\frac{{\bf q}^2}{2 m_\Phi} +  
2\bar{\mu}_\Phi\right)} \right] \right\} \;,  
\label{PPhi}  
\end{eqnarray}  
  
\noindent  
where we have neglected the corrections coming from the new interaction terms
given by the change of the Lagrangian density as in Eq. (\ref{newL}), since
these terms are higher order than the ones considered in the one-loop level.
{}From Eq. (\ref{PPhi}), we determine the total density, $\rho_\Phi$
as\footnote{ Note that here and also for the two-field case discussed below we
  are discarding contributions coming from the derivatives of the self-energy,
  both with relation to the fields and chemical potential, since these terms
  result to be higher order than the one-loop order being considered.  {}For
  instance $\partial \Sigma/\partial \phi_0$ is already of order ${\cal O}
  (g_\Phi^2)$ and same order corrections at the two-loop order should be
  considered for consistency as well.}
  
\begin{eqnarray}  
\rho_\Phi = \frac{\partial P(T,\mu_\Phi)}{\partial \mu_\Phi}  
= \frac{\mu_\Phi}{g_\Phi} - \frac{1}{2} \int \frac {d^3 {\bf q}}{(2 \pi)^3}    
\frac{\frac{{\bf q}^2}{2 m_\Phi}} { \sqrt{ \frac{{\bf q}^2}{2 m_\Phi}  
\left(\frac{{\bf q}^2}{2 m_\Phi} +2\bar{\mu}_\Phi\right)} }  
\left\{ 1+ \frac{2}{\exp\left[\beta \sqrt{ \frac{{\bf q}^2}{2 m_\Phi}  
\left(\frac{{\bf q}^2}{2 m_\Phi} +  
2\bar{\mu}_\Phi\right)} \right] -1 } \right\}\;,  
\label{rhoPhi3}  
\end{eqnarray}  
  
\noindent  
or, also using Eq. (\ref{minimum3}) to express $\mu_\Phi$ in term of
$\bar{\mu}_\Phi$,
  
\begin{eqnarray}  
\rho_\Phi = \frac{\bar{\mu}_\Phi}{g_\Phi} + \frac{1}{2}  
\int \frac {d^3 {\bf q}}{(2 \pi)^3}    
\frac{\frac{{\bf q}^2}{2 m_\Phi} +  
\bar{\mu}_\Phi}{ \sqrt{ \frac{{\bf q}^2}{2 m_\Phi}  
\left(\frac{{\bf q}^2}{2 m_\Phi} +2\bar{\mu}_\Phi\right)} }  
\left\{ 1+ \frac{2}{\exp\left[\beta \sqrt{ \frac{{\bf q}^2}{2 m_\Phi}  
\left(\frac{{\bf q}^2}{2 m_\Phi} +  
2\bar{\mu}_\Phi\right)} \right] -1 } \right\}\;.  
\label{rhoPhi4}  
\end{eqnarray}

\noindent  
The $T=0$ term in the RHS of Eq. (\ref{rhoPhi4}) gives the quantum depletion
of the condensate, while the finite temperature term gives the thermal
depletion, as usual. The $T=0$ momentum integral term can be easily computed
using dimensional regularization.  Converting the momentum integral to
arbitrary $d=3-\epsilon$ dimensions and taking $\epsilon \to 0$, we obtain the
result,
  
\begin{equation}  
\frac{1}{2} \int \frac {d^d {\bf q}}{(2 \pi)^d}    
\frac{\frac{{\bf q}^2}{2 m_\Phi} +  
\bar{\mu}_\Phi}{ \sqrt{ \frac{{\bf q}^2}{2 m_\Phi}  
\left(\frac{{\bf q}^2}{2 m_\Phi} +2\bar{\mu}_\Phi\right)} }  
= \frac{\left( m_\Phi \bar{\mu}_\Phi \right)^{3/2}}{3 \pi^2} +   
{\cal O} (\epsilon)\;.  
\label{T0result}  
\end{equation}

\noindent  
Writing Eq. (\ref{rhoPhi4}) in terms of the temperature dependent condensate
density, $\rho_{\Phi,c}(T)\equiv \phi^2_m(T)/2$ with $\bar{\mu}_\Phi = g_\Phi
\phi_m^2/2$, taking it at the critical point, $T=T_c$, and having that
$\bar{\mu}_\Phi(T=T_c) =0$ (or $\phi_m(T=T_c) =0$, since the condensate
density vanishes at $T_c$), we immediately obtain from Eq. (\ref{rhoPhi4})
that
  
\begin{eqnarray}  
\rho_\Phi = \int \frac {d^3 {\bf q}}{(2 \pi)^3}    
\frac{1}{ e^{ \frac{{\bf q}^2}{2 m_\Phi T_c} }  -1 }  \Rightarrow   
T_{\Phi,c} = \frac{2 \pi}{m_\Phi}  
\left[ \frac{\rho_\Phi}{ \zeta(3/2) } \right]^{2/3}\;,  
\label{rhoPhiTc}  
\end{eqnarray}  
  
\noindent  
where $\zeta(3/2) \simeq 2.612$. The result Eq.(\ref{rhoPhiTc}) is the
standard one for the critical temperature for an homogeneous Bose
gas\footnote{Note that at this level of approximation that we are considering,
  $T_c$ is the same as that for the ideal Bose gas. Corrections due to the
  self-interactions are only accessible through non-perturbative methods
  (beyond one-loop) requiring at least second order corrections in the
  self-energy, see, e.g., Refs. \cite{shift} and \cite{jeans,yuka} for recent
  reviews.}.
  
It is instructive to see, from Eq. (\ref{rhoPhi4}), how the condensate density
$\rho_{\Phi,c}(T)$ changes with temperature.  Writing the coupling constant
$g_\Phi$ in terms of the $s$-wave scattering length $a_\Phi$, $g_\Phi = 4 \pi
a_\Phi/m_\Phi$ and defining the dimensionless quantities
$\tilde{\rho}_{\Phi,c} = \rho_{\Phi,c}/\rho_\Phi$, ${\tilde{T}}_{\Phi} =
T/T_{\Phi,c}$, where $T_{\Phi,c}$ is given by Eq. (\ref{rhoPhiTc}), we obtain
for Eq. (\ref{rhoPhi4}), the implicit equation for
$\tilde{\rho}_{\Phi,c}(\tilde{T})$,
  
\begin{eqnarray}  
1 =  \tilde{\rho}_{\Phi,c} + \frac{8}{3 \pi^{1/2}}  
n_\Phi^{1/2} \tilde{\rho}_{\Phi,c}^{3/2} + \frac{4}{\pi^{1/2}\zeta(3/2)}  
{\tilde{T}}_{\Phi}^{3/2} \int_0^{\infty} dx\, x &&  
\frac{ x^2 + 2 \zeta(3/2)^{2/3} n_\Phi^{1/3}  
\tilde{\rho}_{\Phi,c}/{\tilde{T}}_{\Phi} }{\left[  x^2 +  
4 \zeta(3/2)^{2/3} n_\Phi^{1/3} \tilde{\rho}_{\Phi,c}/{\tilde{T}}_{\Phi}  
\right]^{1/2} } \nonumber \\  
&& \times \frac{1}{e^{x \left[  x^2 +   
4 \zeta(3/2)^{2/3} n_\Phi^{1/3} \tilde{\rho}_{\Phi,c}/{\tilde{T}}_{\Phi}  
\right]^{1/2} } - 1 } \;,  
\label{dless}  
\end{eqnarray}  
  
\noindent  
where we have made the change of integration variable in Eq. (\ref{rhoPhi4}),
$x^2 = q^2/(2 m_\Phi T)$ and also used the quantum depletion result Eq.
(\ref{T0result}). In Eq. (\ref{dless}) we have defined the diluteness
parameter $n_\Phi = \rho_\Phi a_\Phi^3$ \cite{becreview}. The integration in
$x$ in (\ref{dless}) can easily be performed numerically producing the
standard result shown in {}Fig. \ref{phasePhi} for different values of
$n_\Phi$.

\begin{figure}[htb]  
  \vspace{0.5cm}
  \epsfig{figure=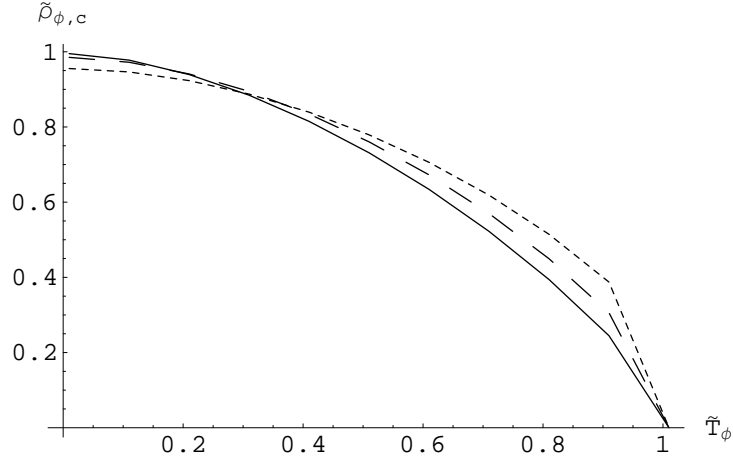,angle=0,width=10cm}
\caption[]{\label{phasePhi} Results for the dimensionless  
  $\tilde{\rho}_{\Phi,c}$ as a function of ${\tilde T}_{\Phi}$, for $n_\Phi=
  10^{-5}$ (solid),$n_\Phi=10^{-4}$ (dashed) and $n_\Phi=10^{-3}$ (dotted).  }
\end{figure}

\subsection{The Two-Field Case}  
  
Let us now return to the two-field case model.  The field propagators in the
non-vanishing self-energy contributions are found by the inverse of the
dressed matrix of quadratic terms, $\hat{M}$, which generalizes Eq.
(\ref{MPhi}) of the one-field case and Eq. (\ref{Mmatrix}) for the free
(inverse of) propagator terms in the $(\phi_1,\phi_2,\psi_1,\psi_2)$ basis. In
momentum space $\hat{M}$ is represented by the matrix
  
\begin{eqnarray}  
\hat{M}_{\Phi,\Psi}(\omega_n,{\bf q}) =    
\left(  
\begin{array}{cccc}  
H_\Phi  + \Sigma_{\phi_1,\phi_1} & -\omega_n  & g \phi_0 \psi_0 +  
\Sigma_{\phi_1,\psi_1} & 0 \\  
\omega_n & G_\Phi  + \Sigma_{\phi_2,\phi_2} & 0 & 0 \\  
g \phi_0 \psi_0 + \Sigma_{\phi_1,\psi_1} & 0 & H_\Psi  + \Sigma_{\psi_1,\psi_1} & -\omega_n \\  
0 & 0 &  \omega_n & G_\Psi  +\Sigma_{\psi_2,\psi_2}  
\end{array}  
\right)\;,  
\label{Mdressed}  
\end{eqnarray}  
  
\noindent  
where the functions $H_i$, $G_i$, $i=\Psi,\Phi$, are given by Eqs.
(\ref{HPsi}) - (\ref{GPhi}), respectively. In obtaining the self-energies is
easy to show that all cross-like self-energies vanish identically except by
$\Sigma_{\phi_1,\psi_1}$, which is given by exchange-like diagrams (like in
{}Fig. 1b).  The introduction of the self-energy terms in (\ref{Mdressed}) can
again be implemented self-consistently already at the Lagrangian level by an
analogous procedure as the one shown in Eq.  (\ref{newL}) for the one-field
case.

In complete analogy with the one-field case we now have, from Eq.
(\ref{Mdressed}), that in order to preserve Goldstone's theorem in both $\Phi$
and $\Psi$ field directions in the broken (condensed) phase, the inclusion of
fluctuations must change the tree-level chemical potentials to
  
\begin{eqnarray}  
&& \bar{\mu}_\Phi = \mu_\Phi - \Sigma_{\phi_2,\phi_2} \;,  
\label{barmuPhi} \\  
&& \bar{\mu}_\Psi = \mu_\Psi - \Sigma_{\psi_2,\psi_2} \;.  
\label{barmuPsi}  
\end{eqnarray}  
  
As in the one-field case, we restrict our main analysis to the high
temperature regime, for which $T \gg {\bar \mu}_i$. In this regime it can
again be checked that the exchange diagrams are subleading compared to the
tadpole diagrams. In this case $\Sigma_{\phi_1,\psi_1}$ can be neglected
compared to the tree level term in Eq. (\ref{Mdressed}) and
$\Sigma_{\phi_1,\phi_1} \simeq \Sigma_{\phi_2,\phi_2}$,
$\Sigma_{\psi_1,\psi_1} \simeq \Sigma_{\psi_2,\psi_2}$. They are given, at the
minima of the effective potential $(\phi_m,\psi_m)$, by
  
\begin{equation}  
\Sigma_{\Phi_i,\Phi_i} \simeq \int \frac {d^3 {\bf q}}{(2 \pi)^3}    
\left[ \frac{\partial \bar{A}}{\partial \phi_0^2}   
\left( 1 + 2 n_{\bar{A}} \right) +  
\frac{\partial \bar{B}}{\partial \phi_0^2}   
\left( 1 + 2 n_{\bar{B}} \right) \right]  
\Bigr|_{ {\lower 1.5pt\hbox{\tiny $\psi_0=\psi_m$}   
\above 0pt \raise 1.5pt\hbox{\tiny $\phi_0=\phi_m$}}} \;,  
\label{SigmaPhi}  
\end{equation}  
and
  
\begin{equation}  
\Sigma_{\Psi_i,\Psi_i} \simeq \int \frac {d^3 {\bf q}}{(2 \pi)^3}    
\left[ \frac{\partial \bar{A}}{\partial \psi_0^2}   
\left( 1 + 2 n_{\bar{A}} \right) +  
\frac{\partial \bar{B}}{\partial \psi_0^2}   
\left( 1 + 2 n_{\bar{B}} \right) \right]  
\Bigr|_{ {\lower 1.5pt\hbox{\tiny $\psi_0=\psi_m$}   
\above 0pt \raise 1.5pt\hbox{\tiny $\phi_0=\phi_m$}}}\;,  
\label{SigmaPsi}  
\end{equation}

\noindent  
with
  
\begin{equation}  
n_{\bar{A}} = \frac{1}{e^{\beta \bar{A} } -1 }\;,\;\;\;\;  
n_{\bar{B}} = \frac{1}{e^{\beta \bar{B} } -1 }\;,  
\label{nAnB}  
\end{equation}  
  
\noindent  
where $\bar{A}$ and $\bar{B}$, which follows from the determinant of Eq.
(\ref{Mdressed}) are the analogous of Eqs. (\ref{A}) and (\ref{B}) with
$\mu_\Phi$ and $\mu_\Psi$ in those expressions replaced by the effective
chemical potentials $\bar{\mu}_\Phi$ and $\bar{\mu}_\Psi$, given by Eqs.
(\ref{barmuPhi}) and (\ref{barmuPsi}), respectively.  The minima of the
effective potential, $\phi_m$ and $\psi_m$, are determined from the effective
potential $V_{\rm eff}(T,\phi_0,\psi_0)$, Eq. (\ref{Veff2}), with $A\to
\bar{A}$ and $B\to \bar{B}$. Note that in this case, just like in the
one-field case, the condensate densities in the presence of fluctuations,
$\rho_{\Psi,c}(T)\equiv \psi_m^2/2$ and $\rho_{\Phi,c} (T) \equiv \phi_m^2/2$,
are given by the same equations (\ref{rhoPsi2}) and (\ref{rhoPhi2}), but with
chemical potentials given by the effective ones,
  
\begin{equation}  
\rho_{\Psi,c}(T)= \frac{g_\Phi \bar{\mu}_\Psi -   
g \bar{\mu}_\Phi}{g_\Psi g_\Phi - g^2} \;,  
\label{rhoPsiT}  
\end{equation}  
and
  
\begin{equation}  
\rho_{\Phi,c}(T)= \frac{g_\Psi \bar{\mu}_\Phi -   
g \bar{\mu}_\Psi}{g_\Psi g_\Phi - g^2} \;.  
\label{rhoPhiT}  
\end{equation}

Let us now compute the total densities, $\rho_\Psi$ and $\rho_\Phi$, which are
obtained from Eq.(\ref{densities}), with the pressure given by
  
\begin{eqnarray}  
P(T,\mu_\Phi, \mu_\Psi) &=&  \frac{1}{ 2 \left( g_\Phi g_\Psi - g^2 \right) }  
\left[ g_\Psi \left( \mu_\Phi^2 -  
\Sigma^2_{\phi_2,\phi_2} \right) + g_\Phi \left( \mu_\Psi^2 -  
\Sigma^2_{\psi_2,\psi_2} \right) + 2 g \left( \Sigma_{\phi_2,\phi_2}  
\Sigma_{\psi_2,\psi_2} - \mu_\Phi \mu_\Psi \right) \right]  
 \nonumber \\  
&-& \int \frac {d^3 {\bf q}}{(2 \pi)^3} \left( \bar{A} + \bar{B} \right)  
\Bigr|_{ {\lower 1.5pt\hbox{\tiny $\psi_0=\psi_m$}   
\above 0pt \raise 1.5pt\hbox{\tiny $\phi_0=\phi_m$}}}  
- \frac{1}{\beta} \int \frac {d^3 {\bf q}}{(2 \pi)^3} \left[  
\ln \left( 1- e^{- \beta \bar{A} } \right)   
+ \ln \left( 1- e^{- \beta \bar{B} } \right) \right]  
\Bigr|_{ {\lower 1.5pt\hbox{\tiny $\psi_0=\psi_m$}   
\above 0pt \raise 1.5pt\hbox{\tiny $\phi_0=\phi_m$}}}\;.  
\label{PPhiPsi}  
\end{eqnarray}  
  
\noindent  
If we now express $\bar{A}$ as $\bar{A} \equiv \bar{A}[
\psi_m(\mu_\Phi,\mu_\Psi,T), \phi_m (\mu_\Phi,\mu_\Psi,T), \mu_\Phi,\mu_\Psi
]$ and $\bar{B}$ as $\bar{B} \equiv \bar{B}[ \psi_m(\mu_\Phi,\mu_\Psi,T),
\phi_m (\mu_\Phi,\mu_\Psi,T), \mu_\Phi,\mu_\Psi ]$ it follows, from Eq.
(\ref{densities}), that we can write for $\rho_\Psi$ and $\rho_\Phi$ the
following expressions:
  
\begin{eqnarray}  
\rho_\Psi &=& \frac{g_\Phi \mu_\Psi - g \mu_\Phi}{g_\Psi g_\Phi - g^2}  
- \int \frac {d^3 {\bf q}}{(2 \pi)^3} \left[   
\frac{g_\Phi}{g_\Psi g_\Phi - g^2} \frac{\partial\bar{A}}  
{\partial \psi_m^2}  
- \frac{g}{g_\Psi g_\Phi - g^2} \frac{\partial \bar{A}}{\partial \phi_m^2}  
+ \frac{1}{2} \frac{\partial \bar{A}}{\partial \mu_\Psi } \right]  
\left( 1 + 2 n_{\bar{A}} \right) \nonumber \\  
&-& \int \frac {d^3 {\bf q}}{(2 \pi)^3} \left[   
\frac{g_\Phi}{g_\Psi g_\Phi - g^2} \frac{\partial\bar{B}}  
{\partial \psi_m^2}  
- \frac{g}{g_\Psi g_\Phi - g^2} \frac{\partial \bar{B}}{\partial \phi_m^2}  
+ \frac{1}{2} \frac{\partial \bar{B}}{\partial \mu_\Psi} \right]  
\left( 1 + 2 n_{\bar{B}} \right) \;,  
\label{rhoPsid}  
\end{eqnarray}  
  
\noindent  
and
  
\begin{eqnarray}  
\rho_\Phi &=& \frac{g_\Psi \mu_\Phi - g \mu_\Psi}{g_\Psi g_\Phi - g^2}  
- \int \frac {d^3 {\bf q}}{(2 \pi)^3} \left[   
\frac{g_\Psi}{g_\Psi g_\Phi - g^2} \frac{\partial\bar{A}}  
{\partial \phi_m^2}  
- \frac{g}{g_\Psi g_\Phi - g^2} \frac{\partial \bar{A}}{\partial \psi_m^2}  
+ \frac{1}{2} \frac{\partial \bar{A}}{\partial \mu_\Phi} \right]  
\left( 1 + 2 n_{\bar{A}} \right) \nonumber \\  
&-& \int \frac {d^3 {\bf q}}{(2 \pi)^3} \left[   
\frac{g_\Psi}{g_\Psi g_\Phi - g^2} \frac{\partial\bar{B}}  
{\partial \phi_m^2}  
- \frac{g}{g_\Psi g_\Phi - g^2} \frac{\partial \bar{B}}{\partial \psi_m^2}  
+ \frac{1}{2} \frac{\partial \bar{B}}{\partial \mu_\Phi } \right]  
\left( 1 + 2 n_{\bar{B}} \right) \;.  
\label{rhoPhid}  
\end{eqnarray}  
  
\noindent  
We now make use of the expressions for the condensate densities at finite
temperature, Eqs.  (\ref{rhoPsiT}) and (\ref{rhoPhiT}), with Eqs.
(\ref{barmuPhi}) and (\ref{barmuPsi}) together with the self-energies
expressions Eqs. (\ref{SigmaPhi}) and (\ref{SigmaPsi}), to express Eqs.
(\ref{rhoPsid}) and (\ref{rhoPhid}) completely in terms of the temperature
dependent condensate densities (e.g., in terms of $\phi_m$ and $\psi_m$)
instead of the tree-level ($T=0$) condensate densities $\phi_0$ and $\psi_0$.
This process is analogous to the one used to obtain Eq. (\ref{rhoPhi4}) for
the one-field case.  After some straightforward algebra, this then results in
the coupled equations expressing $\psi_m$ and $\phi_m$ in terms the total
densities $\rho_\Psi$ and $\rho_\Phi$,
  
\begin{equation}  
\rho_\Psi = \frac{\psi_m^2}{2} - \frac{1}{2} 
\int \frac {d^3 {\bf q}}{(2 \pi)^3}  
\left[ \frac{\partial\bar{A}} {\partial \mu_\Psi} 
\left( 1 + 2 n_{\bar{A}} \right)  
+ \frac{\partial\bar{B}} {\partial \mu_\Psi} 
\left( 1 + 2 n_{\bar{B}} \right)  
\right] \;,  
\label{rhopsid2}  
\end{equation}  
  
\noindent  
and
  
\begin{equation}  
\rho_\Phi = \frac{\phi_m^2}{2} - \frac{1}{2} 
\int \frac {d^3 {\bf q}}{(2 \pi)^3}  
\left[ \frac{\partial\bar{A}} {\partial \mu_\Phi} 
\left( 1 + 2 n_{\bar{A}} \right)  
+ \frac{\partial\bar{B}} {\partial \mu_\Phi} 
\left( 1 + 2 n_{\bar{B}} \right)  
\right] \;,  
\label{rhophid2}  
\end{equation}  
  
\noindent  
Using Eqs. (\ref{A}) and (\ref{B}, with $\mu_i \to \bar{\mu}_i$, we obtain
that

\begin{eqnarray}  
&& \frac{\partial \bar{A}}{\partial \mu_\Phi} =  
-\frac{1}{4 \bar{A} } \left( \bar{H}_\Phi + \bar{G}_\Phi \right)  
- \frac{1}{4 \bar{A} } \frac{ \left( \bar{H}_\Phi + \bar{G}_\Phi \right)  
\left( \bar{H}_\Psi \bar{G}_\Psi - \bar{H}_\Phi \bar{G}_\Phi \right)  
- 2 g^2 \phi_m^2 \psi_m^2 \bar{G}_\Psi }{  
\sqrt{ \left( \bar{H}_\Psi \bar{G}_\Psi - \bar{H}_\Phi \bar{G}_\Phi \right)^2  
+ 4 g^2 \phi_m^2 \psi_m^2 \bar{G}_\Phi \bar{G}_\Psi} } \;,   
\nonumber \\  
&& \frac{\partial \bar{A}}{\partial \mu_\Psi} =  
-\frac{1}{4 \bar{A} } \left( \bar{H}_\Psi + \bar{G}_\Psi \right)  
+ \frac{1}{4 \bar{A} } \frac{ \left( \bar{H}_\Psi + \bar{G}_\Psi \right)  
\left( \bar{H}_\Psi \bar{G}_\Psi - \bar{H}_\Phi \bar{G}_\Phi \right)  
+ 2 g^2 \phi_m^2 \psi_m^2 \bar{G}_\Phi }{  
\sqrt{ \left( \bar{H}_\Psi \bar{G}_\Psi - \bar{H}_\Phi \bar{G}_\Phi \right)^2  
+ 4 g^2 \phi_m^2 \psi_m^2 \bar{G}_\Phi \bar{G}_\Psi} } \;,   
\nonumber \\  
&& \frac{\partial \bar{B}}{\partial \mu_\Phi} =  
-\frac{1}{4 \bar{B} } \left( \bar{H}_\Phi + \bar{G}_\Phi \right)  
+ \frac{1}{4 \bar{B} } \frac{ \left( \bar{H}_\Phi + \bar{G}_\Phi \right)  
\left( \bar{H}_\Psi \bar{G}_\Psi - \bar{H}_\Phi \bar{G}_\Phi \right)  
- 2 g^2 \phi_m^2 \psi_m^2 \bar{G}_\Psi }{  
\sqrt{ \left( \bar{H}_\Psi \bar{G}_\Psi - \bar{H}_\Phi \bar{G}_\Phi \right)^2  
+ 4 g^2 \phi_m^2 \psi_m^2 \bar{G}_\Phi \bar{G}_\Psi} } \;,   
\nonumber \\  
&& \frac{\partial \bar{B}}{\partial \mu_\Psi} =  
-\frac{1}{4 \bar{B} } \left( \bar{H}_\Psi + \bar{G}_\Psi \right)  
- \frac{1}{4 \bar{B} } \frac{ \left( \bar{H}_\Psi + \bar{G}_\Psi \right)  
\left( \bar{H}_\Psi \bar{G}_\Psi - \bar{H}_\Phi \bar{G}_\Phi \right)  
+ 2 g^2 \phi_m^2 \psi_m^2 \bar{G}_\Phi }{  
\sqrt{ \left( \bar{H}_\Psi \bar{G}_\Psi - \bar{H}_\Phi \bar{G}_\Phi \right)^2  
+ 4 g^2 \phi_m^2 \psi_m^2 \bar{G}_\Phi \bar{G}_\Psi} } \;,   
\label{partialAB}  
\end{eqnarray}  
  
\noindent  
with
  
\begin{eqnarray}  
\bar{H}_\Psi &=&  \frac{{\bf q}^2}{2 m_\Psi} + g_\Psi \psi_m^2\;, \nonumber \\  
\bar{G}_\Psi &=&  \frac{{\bf q}^2}{2 m_\Psi} \;, \nonumber \\  
\bar{H}_\Phi &=&  \frac{{\bf q}^2}{2 m_\Phi} + g_\Phi \phi_m^2\;, \nonumber \\  
\bar{G}_\Phi &=&  \frac{{\bf q}^2}{2 m_\Phi} \;.  
\label{HG}  
\end{eqnarray}  
  
The coupled equations (\ref{rhopsid2}) and (\ref{rhophid2}) seem very messy.
But still we can obtain a few analytical results from it and perform some
qualitative discussions about the phase diagram $(\rho_\Phi,\rho_\Psi, T)$.
for the coupled system. {}For instance, if we consider an equal mass system,
$m_\Phi = m_\Psi =m$, the expressions for $\bar{A}$ and $\bar{B}$ simplify to
  
\begin{eqnarray}  
&&\bar{A}^2 =   \frac{{\bf q}^2}{2 m} \left(\frac{{\bf q}^2}{2 m}+  
\alpha_-\right) \;, \nonumber \\  
&&\bar{B}^2 =   \frac{{\bf q}^2}{2 m} \left(\frac{{\bf q}^2}{2 m}+  
\alpha_+\right) \;,  
\label{ABm}  
\end{eqnarray}  
  
\noindent  
with
  
\begin{equation}  
\alpha_\pm = \frac{g_\Psi \psi_m^2}{2} +\frac{g_\Phi \phi_m^2}{2}  
\pm \left[ \left(  \frac{g_\Psi \psi_m^2}{2} -\frac{g_\Phi \phi_m^2}{2}  
\right)^2 + g^2 \psi_m^2 \phi_m^2 \right]^{1/2}\;.  
\label{alpha+-}  
\end{equation}  
  
\noindent  
Using Eqs. (\ref{ABm}) and (\ref{alpha+-}) in Eqs. (\ref{rhopsid2}) and
(\ref{rhophid2}) we can compute the quantum depletion terms (the $T=0$ terms)
appearing in the coupled system of equations, Eqs. (\ref{rhopsid2}) and
(\ref{rhophid2}).  Like in the one field case, we use again dimensional
regularization to compute the zero temperature momentum integrals in Eqs.
(\ref{rhopsid2}) and (\ref{rhophid2}) to obtain the results,
  
\begin{eqnarray}  
- \frac{1}{2} \int \frac {d^d {\bf q}}{(2 \pi)^d} \left(  
\frac{\partial\bar{A}} {\partial \mu_\Psi}   
+ \frac{\partial\bar{B}} {\partial \mu_\Psi} \right)  
&=& \frac{(2 m \alpha_+)^{3/2} + (2 m \alpha_-)^{3/2}}{12 \pi^2}  
\nonumber \\  
&+& \frac{\left [\frac{g_\Psi \psi_m^2}{2} -\frac{g_\Phi \phi_m^2 }{2} \right] } {\left[ \left(  
\frac{g_\Psi \psi_m^2}{2} -\frac{g_\Phi \phi_m^2}{2}  
\right)^2 + g^2 \psi_m^2 \phi_m^2 \right]^{1/2}} \frac{(2 m \alpha_+)^{3/2} -  
(2 m \alpha_-)^{3/2}}{12 \pi^2}  
\nonumber \\  
&-& \left (\frac{m g_\Psi \psi_m^2}{8 \pi^2} \right )\left[  
(2 m \alpha_+)^{1/2} + (2 m \alpha_-)^{1/2}\right]  
\nonumber \\  
&-& \left (\frac{m}{4 \pi^2}\right ) \frac{ \left [g_\Psi \frac{\psi_m^2}{2} \left(  
\frac{g_\Psi \psi_m^2}{2} - \frac{g_\Phi \phi_m^2}{2} \right)+  
g^2 \psi_m^2 \phi_m^2\right ]}{\left[ \left(  
\frac{g_\Psi \psi_m^2}{2} -\frac{g_\Phi \phi_m^2}{2}  
\right)^2 + g^2 \psi_m^2 \phi_m^2 \right]^{1/2}}  
\left[ (2 m \alpha_+)^{1/2} - (2 m \alpha_-)^{1/2}\right]\;,  
\label{int1}  
\end{eqnarray}  
  
\noindent  
and
  
\begin{eqnarray}  
- \frac{1}{2} \int \frac {d^d {\bf q}}{(2 \pi)^d} \left(  
\frac{\partial\bar{A}} {\partial \mu_\Phi}   
+ \frac{\partial\bar{B}} {\partial \mu_\Phi} \right)  
&=& \frac{(2 m \alpha_+)^{3/2} + (2 m \alpha_-)^{3/2}}{12 \pi^2}  
\nonumber \\  
&-& \frac{\left[ \frac{g_\Psi \psi_m^2}{2} -\frac{g_\Phi \phi_m^2}{2}\right ]}{\left[ \left(  
\frac{g_\Psi \psi_m^2}{2} -\frac{g_\Phi \phi_m^2}{2}  
\right)^2 + g^2 \psi_m^2 \phi_m^2 \right]^{1/2}} \frac{(2 m \alpha_+)^{3/2} -  
(2 m \alpha_-)^{3/2}}{12 \pi^2}  
\nonumber \\  
&-& \left (\frac{m g_\Phi \phi_m^2}{8 \pi^2} \right )\left[  
(2 m \alpha_+)^{1/2} + (2 m \alpha_-)^{1/2}\right]  
\nonumber \\  
&+& \left ( \frac{m}{4 \pi^2} \right )\frac{\left [ g_\Phi \frac{\phi_m^2}{2} \left(  
\frac{g_\Psi \psi_m^2}{2} - \frac{g_\Phi \phi_m^2}{2} \right)+  
g^2 \psi_m^2 \phi_m^2 \right ]}{\left[ \left(  
\frac{g_\Psi \psi_m^2}{2} -\frac{g_\Phi \phi_m^2}{2}  
\right)^2 + g^2 \psi_m^2 \phi_m^2 \right]^{1/2}}  
\left[ (2 m \alpha_+)^{1/2} - (2 m \alpha_-)^{1/2}\right]\;.  
\label{int2}  
\end{eqnarray}  
  
\noindent  
{}From either Eq. (\ref{int1}) or Eq. (\ref{int2}), it can easily be checked
that for $g=0$ we re-obtain the result (\ref{T0result}) for either the $\Psi$
or the $\Phi$ fields.  Note also that by taking $g=0$ Eqs.  (\ref{rhopsid2})
and (\ref{rhophid2}) decouple and we recover the one-field expression
(\ref{rhoPhi4}) for each of the fields individually. It is interesting to
point out that, at this level of approximation we are considering, the above
equations show that if any of the fields go above the transition point (either
$\phi_m=0$ or $\psi_m=0$) the two equations (\ref{rhopsid2}) and
(\ref{rhophid2}) decouple, becoming independent of each other, since the
cross-coupling term in (\ref{partialAB}) always appears multiplying both
$\phi_m$ and $\psi_m$.  As far as SNR/ISB are concerned one may conclude,
based on the above equations, that these phenomena do not arise for this
theory since the cross coupling always appears as $g^2$ so the relevant
physical quantities are insensitive to the sign of $g$.

Equations (\ref{rhopsid2}) and (\ref{rhophid2}) are the two-field analogous of
the one-field case, Eq. (\ref{rhoPhi4}). Note that here it is more convenient
to express the resulting expressions for the densities completely in terms of
$\phi_m$ and $\psi_m$. In the one-field case both the condensate density and
the effective chemical potential, for definition, vanish at the critical
point. In the two-field case, as Eqs. (\ref{rhoPsiT}) and (\ref{rhoPhiT})
show, at the critical points for $\Phi$ and $\Psi$ we have $\phi_m(T=T_{\Phi})
=0$ and $\psi_m(T=T_{\Psi}) =0$. However, the same does not necessarily (as it
should not, actually) happens with the effective chemical potentials,
$\bar{\mu}_\Phi$ and $\bar{\mu}_\Psi$. The critical temperature for
transitions in the $\Phi$ and $\Psi$ directions are to be determined from the
numerical solution of the coupled equations (\ref{rhopsid2}) and
(\ref{rhophid2}). {}For this, it is useful to express Eqs. (\ref {rhopsid2})
and (\ref {rhophid2}) into dimensionless quantities, like in Eq. (\ref{dless})
for the one-field case. We start by making the definitions $\rho_\Psi = \theta
\rho_\Phi$, $m_\Psi=m_\Phi=m$, $a_{\Psi,\Phi}^2 = \gamma a_\Psi a_\Phi$ and
$a_i = (n_i/\rho_i)^{1/3}$ where $n_i$ ($i= \Psi,\Phi$) are the diluteness
parameters for $\Psi$ and $\Phi$. It then follows that
$g_i=(4\pi/m)(n_i/\rho_i)^{1/3}$, $g^2= \gamma (8\pi/m)^2
(n_\Psi/\rho_\Psi)^{1/3} (n_\Phi/\rho_\Phi)^{1/3}$, whereas the boundness
condition now reads just $\gamma < 1/4$. As in the monoatomic case,
$x^2=q^2/(2mT)$.  Other useful quantities are the dimensionless temperatures
${\tilde T}_i = T/T_{i,c}$, where $T_{i,c}= (2\pi/m)(\rho_i/\zeta(3/2))^{2/3}$
represents the critical temperature for the monoatomic case, and the
dimensionless integral measure, $(1/\rho_i)[d^3 q/(2\pi)^3] =
(1/\zeta(3/2))(4/\sqrt{\pi}) {\tilde T}_i^{3/2} x^2 dx$. Then, Eqs.
(\ref{rhopsid2}) and ( \ref {rhophid2}) can be written as
 
\begin{equation}  
1= {\tilde \rho}_{c,i} - \frac{1}{2 \rho_i} \int \frac{d^3 {\bf q}}{(2\pi)^3}  
\left ( \frac{ \partial \bar{A}}{\partial \mu_i} + 
\frac{ \partial \bar{B}}{\partial \mu_i} \right ) -  
\frac{1}{ \rho_i} \int \frac{d^3 {\bf q}}{(2\pi)^3}  
\left ( \frac{ \partial \bar{A}}{\partial \mu_i}n_{\bar A} + 
\frac{ \partial \bar{B}}{\partial \mu_i}n_{\bar B} \right ) \;, 
\end{equation}  
where the temperature independent (quantum depletion) part is given by
 
\begin{eqnarray}  
- \frac{1}{2 \rho_\Psi} \int \frac{d^3 {\bf q}}{(2\pi)^3}  
\left ( \frac{ \partial \bar{A}}{\partial \mu_\Psi} + 
\frac{ \partial \bar{B}}{\partial \mu_\Psi} \right )  
&=& \frac {1}{12\pi^2} [ (2{\tilde \alpha}_+)^{3/2}+  
(2{\tilde \alpha}_{-})^{3/2}] \nonumber \\  
&+& \frac{1}{12\pi^2} \frac { \left[ n_\Psi^{1/3} {\tilde \rho}_{\Psi,c} -  
n_\Phi^{1/3} {\tilde \rho}_{\Phi,c}/\theta^{2/3} \right ] 
[ (2{\tilde \alpha}_+)^{3/2}- (2{\tilde \alpha}_{-})^{3/2}]}  
{ \left \{\left[ n_\Psi^{1/3} {\tilde \rho}_{\Psi,c} -  
n_\Phi^{1/3} {\tilde \rho}_{\Phi,c}/\theta^{2/3} \right ]^2 +  
16 (\gamma/\theta^{2/3})  
n_\Psi^{1/3}n_\Phi^{1/3}{\tilde \rho}_{\Psi,c} 
{\tilde \rho}_{\Phi,c} \right \}^{1/2} }  
\nonumber \\  
&-& \frac{1}{\pi}(n_\Psi^{1/3}{\tilde \rho}_{\Psi,c}) 
[ (2{\tilde \alpha}_+)^{1/2}+ (2{\tilde \alpha}_{-})^{1/2}] \nonumber \\  
&-& \frac{1}{\pi} \frac{[(n_\Psi^{1/3}{\tilde \rho}_{\Psi,c}) 
( n_\Psi^{1/3} {\tilde \rho}_{\Psi,c} -  
n_\Phi^{1/3} {\tilde \rho}_{\Phi,c}/\theta^{2/3}) +  
16 (\gamma/\theta^{2/3})  
n_\Psi^{1/3}n_\Phi^{1/3}{\tilde \rho}_{\Psi,c}{\tilde \rho}_{\Phi,c})]}{  
\left \{[ n_\Psi^{1/3} {\tilde \rho}_{\Psi,c} -  
n_\Phi^{1/3} {\tilde \rho}_{\Phi,c}/\theta^{2/3}]^2 +  
16 (\gamma/\theta^{2/3})  
n_\Psi^{1/3}n_\Phi^{1/3}{\tilde \rho}_{\Psi,c}{\tilde \rho}_{\Phi,c}) \right \}^{1/2}}  
\nonumber \\  
&\times&[ (2{\tilde \alpha}_+)^{1/2}- (2{\tilde \alpha}_{-})^{1/2}] \;, 
\end{eqnarray}  
where
 
\begin{equation}  
{\tilde \alpha}_\pm=(4\pi)\left \{n_\Psi^{1/3} {\tilde \rho}_{\Psi,c} +  
n_\Phi^{1/3} \frac{{\tilde \rho}_{\Phi,c}}{\theta^{2/3}} \pm \left [  
\left (n_\Psi^{1/3} {\tilde \rho}_{\Psi,c} -  
n_\Phi^{1/3} \frac{{\tilde \rho}_{\Phi,c}}{\theta^{2/3}}\right )^2  
+ 16 \frac{\gamma}{\theta^{2/3}}  
n_\Psi^{1/3}n_\Phi^{1/3}{\tilde \rho}_{\Psi,c}{\tilde \rho}_{\Phi,c} \right ]^{1/2}  
\right \} \;. 
\end{equation}  
Then, using $\rho_\Phi=\rho_\Psi/\theta$ one gets
  
\begin{eqnarray}  
- \frac{1}{2 \rho_\Phi} \int \frac{d^3 {\bf q} }{(2\pi)^3}  
\left ( \frac{ \partial \bar{A}}{\partial \mu_\Phi} +\frac{ \partial \bar{B}}{\partial \mu_\Phi} \right )  
&=& \frac {\theta}{12\pi^2} [ (2{\tilde \alpha}_+)^{3/2}+  
(2{\tilde \alpha}_{-})^{3/2}] \nonumber \\  
&-& \frac{\theta}{12\pi^2} \frac { \left[ n_\Psi^{1/3}  
{\tilde \rho}_{\Psi,c} -  
n_\Phi^{1/3} {\tilde \rho}_{\Phi,c}/\theta^{2/3} \right ] 
[ (2{\tilde \alpha}_+)^{3/2}- (2{\tilde \alpha}_{-})^{3/2}]}  
{ \left \{\left[ n_\Psi^{1/3} {\tilde \rho}_{\Psi,c} -  
n_\Phi^{1/3} {\tilde \rho}_{\Phi,c}/\theta^{2/3} \right ]^2  
+ 16 (\gamma/\theta^{2/3})  
n_\Psi^{1/3}n_\Phi^{1/3}{\tilde \rho}_{\Psi,c}{\tilde \rho}_{\Phi,c}  
\right \}^{1/2} }  
\nonumber \\  
&-& \frac{\theta^{1/3}}{\pi}(n_\Phi^{1/3}{\tilde \rho}_{\Phi,c}) 
[ (2{\tilde \alpha}_+)^{1/2}+ (2{\tilde \alpha}_{-})^{1/2}] \nonumber \\  
&+& \frac{\theta^{1/3}}{\pi} \frac{[(n_\Phi^{1/3}{\tilde \rho}_{\Phi,c}) 
( n_\Psi^{1/3} {\tilde \rho}_{\Psi,c} -  
n_\Phi^{1/3} {\tilde \rho}_{\Phi,c}/\theta^{2/3}) + 16 \gamma  
n_\Psi^{1/3}n_\Phi^{1/3}{\tilde \rho}_{\Psi,c}{\tilde \rho}_{\Phi,c})]}{  
\left \{[ n_\Psi^{1/3} {\tilde \rho}_{\Psi,c} -  
n_\Phi^{1/3} {\tilde \rho}_{\Phi,c}/\theta^{2/3}]^2 + 16 (\gamma/\theta^{2/3})  
n_\Psi^{1/3}n_\Phi^{1/3}{\tilde \rho}_{\Psi,c}{\tilde \rho}_{\Phi,c})  
\right \}^{1/2}}  
\nonumber \\  
&\times&[ (2{\tilde \alpha}_+)^{1/2}- (2{\tilde \alpha}_{-})^{1/2}] \;. 
\end{eqnarray}  
Let us now use ${\tilde T}_\Psi= T/T_{\Psi,c}$ as a reference temperature to
define the temperature dependent parts, starting with
 
\begin{equation}  
-\frac{1}{ \rho_\Psi} \int \frac{d^3 {\bf q}}{(2\pi)^3}  
\left ( \frac{ \partial A}{\partial \mu_\Psi}n_A +\frac{ \partial B}{\partial \mu_\Psi}n_B \right )  
=- \frac {4}{\sqrt{\pi} \zeta(3/2)} {\tilde T}_\Psi^{3/2}  
\int_0^{\infty}  
dx\, x^2  
\left [ \frac{ \partial A(x)}{\partial \mu_\Psi}n_{A(x)} + 
\frac{ \partial B(x)}{\partial \mu_\Psi}n_{ B(x)} \right ] \;, 
\end{equation}  
where the dimensionless Bose factors are $n_{A(x)} = 1/(e^{A(x)} -1)$ and
$n_{B(x)} = 1/(e^{B(x)} -1)$, with
 
\begin{equation}  
{A(x)} = x \left [ x^2 + \left (\frac {{\tilde \alpha}_- 
\zeta(3/2)^{2/3}}{2\pi{\tilde T}_\Psi}\right ) \right ]^{1/2} \;, 
\end{equation}  
and
 
\begin{equation}  
{B(x)} = x \left [ x^2 + \left (\frac {{\tilde \alpha_+} 
\zeta(3/2)^{2/3}}{2\pi{\tilde T}_\Psi}\right ) \right ]^{1/2} \;. 
\end{equation}  
Let us further define
 
\begin{equation}  
{\cal F}_\Psi(x) = \left [ x^2 + 2 \zeta(3/2)^{2/3}n_\Psi^{1/3}  
\frac{{\tilde \rho}_{\Psi,c}}{{\tilde T}_\Psi} \right ] \;, 
\end{equation}  
  
\begin{equation}  
{\cal G}(x) = x^2\left [ x^2 + 4 \zeta(3/2)^{2/3}n_\Psi^{1/3}  
\frac{{\tilde \rho}_{\Psi,c}}{{\tilde T}_\Psi}\right ] -  
x^2\left [ x^2 + 4 \left( \frac{\zeta(3/2)}{\theta} \right)^{2/3}  
n_\Phi^{1/3} \frac{{\tilde \rho}_{\Phi,c}}{{\tilde T}_\Psi}\right ] \;, 
\end{equation}  
and
 
\begin{equation}  
{\cal H}(x) = \left [ {\cal G}(x)^2 +64 \gamma  
\frac{\zeta(3/2)^{4/3}}{\theta^{2/3}}  
n_\Psi^{1/3}n_\Phi^{1/3}{\tilde \rho}_{\Psi,c}{\tilde \rho}_{\Phi,c}  
\frac {x^4}{{\tilde T}_\Psi^2} \right ]^{1/2} \,\,.  
\end{equation}  
Then,
 
\begin{equation}  
\frac{\partial \bar{A}}{\mu_\Psi} = -\frac{1}{2{A(x)}} 
\left [ {\cal F}_\Psi(x) - \frac{1}{{\cal H}(x)} 
\left ({\cal F}_\Psi(x){\cal G}(x)+16 \gamma  
\frac{\zeta(3/2)^{4/3}}{\theta^{2/3}}  
n_\Psi^{1/3}n_\Phi^{1/3}{\tilde \rho}_{\Psi,c}{\tilde \rho}_{\Phi,c}  
\frac{x^2}{{\tilde T}_\Psi^2}\right ) \right ] \;, 
\end{equation}  
and
 
\begin{equation}  
\frac{\partial \bar{B}}{\mu_\Psi} = -\frac{1}{2{ B(x)}} 
\left [ {\cal F}_\Psi(x) + \frac{1}{{\cal H}(x)} 
\left ({\cal F}_\Psi(x){\cal G}(x)+16 \gamma  
\frac{\zeta(3/2)^{4/3}}{\theta^{2/3}}  
n_\Psi^{1/3}n_\Phi^{1/3}{\tilde \rho}_{\Psi,c}{\tilde \rho}_{\Phi,c}  
\frac{x^2}{{\tilde T}_\Psi^2}\right ) \right ] \;. 
\end{equation}  
{}Finally, let us write down
 
\begin{equation}  
-\frac{1}{ \rho_\Phi} \int \frac{d^3 {\bf q}}{(2\pi)^3}  
\left ( \frac{ \partial \bar{A}}{\partial \mu_\Phi}n_{\bar A} + 
\frac{ \partial \bar{B}}{\partial \mu_\Phi}n_{\bar B} \right )  
=- \frac {4}{\sqrt{\pi} \zeta(3/2)} {\tilde T}_\Psi^{3/2} \theta 
\int_0^{\infty} dx \, x^2  
\left ( \frac{ \partial A(x)}{\partial \mu_\Phi}n_{A(x)} + 
\frac{ \partial B(x)}{\partial \mu_\Phi}n_{ B(x)} \right ) \,\,,  
\end{equation}  
using
 
\begin{equation}  
{\cal F}_\Phi(x) = \left [x^2 + 2 \zeta(3/2)^{2/3}n_\Phi^{1/3}  
\frac{{\tilde \rho}_{\Phi,c}}{\theta^{2/3}{\tilde T}_\Psi} \right ] \,\,,  
\end{equation}  
and the above definitions for ${\cal G}$ and ${\cal H}$.  Then,
 
\begin{equation}  
\frac{\partial \bar{A}}{\mu_\Phi} = -\frac{1}{2{ A(x)}} 
\left [ {\cal F}_\Phi(x) + \frac{1}{{\cal H}(x)} 
\left ({\cal F}_\Phi(x){\cal G}(x)-16 \gamma  
\frac{\zeta(3/2)^{4/3}}{\theta^{2/3}}  
n_\Psi^{1/3}n_\Phi^{1/3}{\tilde \rho}_{\Psi,c}{\tilde \rho}_{\Phi,c}  
\frac{x^2}{{\tilde T}_\Psi^2}\right ) \right ] \;, 
\end{equation}  
and
 
\begin{equation}  
\frac{\partial \bar{B}}{\mu_\Phi} = -\frac{1}{2{B(x)}} 
\left [ {\cal F}_\Phi(x) - \frac{1}{{\cal H}(x)} 
\left ({\cal F}_\Phi(x){\cal G}(x)-16 \gamma  
\frac{\zeta(3/2)^{4/3}}{\theta^{2/3}}  
n_\Psi^{1/3}n_\Phi^{1/3}{\tilde \rho}_{\Psi,c}{\tilde \rho}_{\Phi,c}  
\frac{x^2}{{\tilde T}_\Psi^2}\right ) \right ] \;. 
\end{equation}

We have carried out a careful numerical analysis finding that the cross
coupling (now characterized by the dimensionless parameter $\gamma$) does not
affect the critical temperature that each gas observes when the system is
uncoupled. As shown in {} Fig. \ref {allrhodif}, when the temperature is
increased each type of condensate returns to the symmetric (gas) phase at a
critical temperature whose value coincides with the one obtained in the one
field case (see Eq. (\ref {rhoPhiTc})). That is, the two distinct critical
temperatures displayed in {} Fig. \ref {allrhodif} are insensitive to $\gamma$
({\it i.e.} $g$) and seem to depend separately on each density \footnote{
  Actually, it is very plausible that the critical temperatures should also
  depend on the different masses as in Eq.(\ref {rhoPhiTc}). However, here we
  had to use the approximation $m_\Phi \simeq m_\Psi=m$ which does not allow
  us to fully confirm this fact.} that are now related by $\theta$. Although
the critical temperatures for each gas in the coupled case coincide with the
ones observed in the monoatomic case the temperature dependence of the
quantities ${\tilde \rho}_{c,i}$ are influenced by the cross-coupling.
 
\begin{figure}[htb]  
  \vspace{0.5cm}
  \epsfig{figure=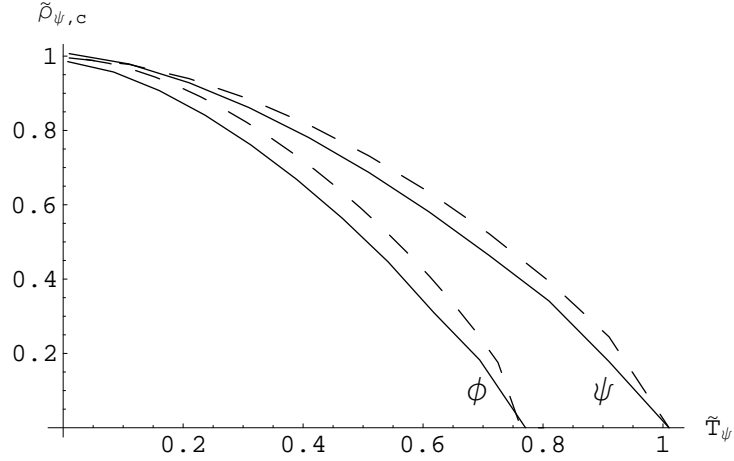,angle=0,width=10cm}
\caption[]{\label{allrhodif} The  quantities  ${\tilde \rho}_{c,\Psi}$ and   
  ${\tilde \rho}_{c,\Phi}$ plotted as functions of the reference dimensionless
  temperature ${\tilde T}_{\Psi}$ for $\gamma=1/6$, $\theta = 1.5$ and
  $n_\Psi=n_\Phi=10^{-5}$ The solid lines correspond to the coupled case
  whereas the dashed lines are the corresponding curves for the one field
  case. Note how the observed $T_c$'s agree with the value predicted for the
  one field case, Eq. (\ref {rhoPhiTc}). The symmetry $O(2)$ referring to the
  less dense gas, $\Phi$, is restored at a lower temperature ${\tilde
    T}_{\Psi}\simeq 0.763$ (which corresponds to ${\tilde T}_{\Phi}= \theta
  {\tilde T}_{\Psi}=1$) than the one observed by the other specie, $\Psi$
  (${\tilde T}_{\Psi}=1$). }
\end{figure}

To investigate how the cross coupling influences the general temperature
behavior of ${\tilde \rho}_{c,i}$, we consider a large, a small and the null
value for $\gamma$. {}Figure \ref {difgama} shows the situation for the $\Psi$
sector displaying the fact that, for $0< {\tilde T}_{\Psi} < 1$, the density
${\tilde \rho}_{c,\Psi}$ assumes smaller values for larger values of the cross
coupling. However, as noticed before, the critical value for the temperature
corresponds to the monoatomic case that in fact is the critical temperature
for a non interacting gas. This is expected since at this (one-loop) level of
approximation the Hugenholtz-Pines theorem washes out all nontrivial
contributions to $T_c$ (see Ref. \cite {shift}) so that our results for this
quantity become trivial. Nevertheless {}Fig. \ref {difgama} suggests that the
critical temperature value may be influenced by the cross coupling interaction
in a computation that includes higher corrections.
  
\begin{figure}[htb]  
  \vspace{0.5cm}
  \epsfig{figure=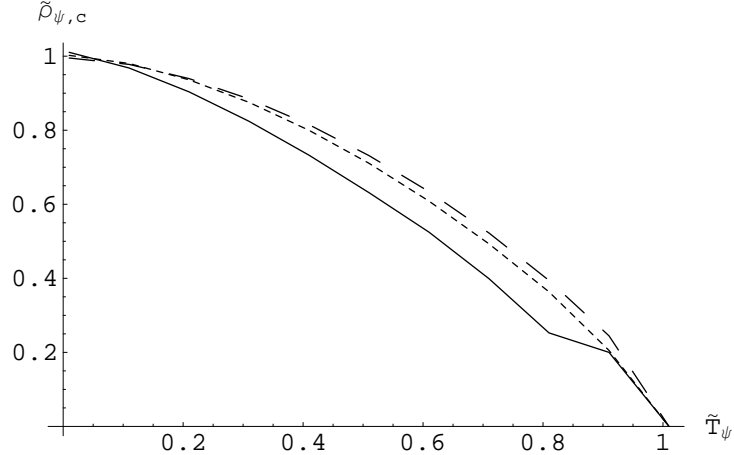,angle=0,width=10cm}
\caption[]{\label{difgama} The  quantity  ${\tilde \rho}_{c,\Psi}$  as a  
  function of the dimensionless temperature ${\tilde T}_{\Psi}$ for $\theta =
  1.0$ and $n_\Psi=n_\Phi=10^{-5}$. The solid line corresponds to a strongly
  coupled case, $\gamma =2/9$, whereas the dotted line corresponds to a weakly
  coupled case, $\gamma= 1/10$. The dashed line corresponds to the uncoupled
  case.}
\end{figure}

\section{The Symmetry Restored, High Temperature Phases}  
  
Lets us now study the two-coupled field system in the symmetry restored phase.
In this case the spectrum for both $\Phi$ and $\Psi$ field changes, since
$\phi_m$ and $\psi_m$ both vanish. The functions $H_\Phi$ and $G_\Phi$ defined
by Eqs. (\ref{HPhi2}) and (\ref{GPhi2}), and similarly for $H_\Psi$ and
$G_\Psi$, become
  
\begin{eqnarray}  
H_\Phi = G_\Phi &=&   \frac{{\bf q}^2}{2 m_\Phi} - \bar{\mu}_\Phi \;,   
\label{HGPhi}\\  
H_\Psi = G_\Psi &=&   \frac{{\bf q}^2}{2 m_\Psi} - \bar{\mu}_\Psi\;,   
\label{HGPsi}  
\end{eqnarray}

\noindent  
where $\bar{\mu}_\Phi = \mu_\Phi - \Sigma_{\Phi,\Phi}$ and $\bar{\mu}_\Psi =
\mu_\Psi - \Sigma_{\Psi,\Psi}$, with the self-energies $\Sigma$, in the normal
phase and at one-loop order, given by
  
\begin{eqnarray}  
\Sigma_{\Phi,\Phi} \equiv \Sigma_{\phi_1,\phi_1}=\Sigma_{\phi_2,\phi_2} &=&   
g_\Phi \int \frac {d^3 {\bf q}}{(2 \pi)^3}    
\left\{ 1+ \frac{2}{\exp\left[\beta   
\left(\frac{{\bf q}^2}{2 m_\Phi} -  
\bar{\mu}_\Phi\right) \right] -1 } \right\} \nonumber \\  
&+& \frac{g}{2} \int \frac {d^3 {\bf q}}{(2 \pi)^3}    
\left\{ 1+ \frac{2}{\exp\left[\beta   
\left(\frac{{\bf q}^2}{2 m_\Psi} -  
\bar{\mu}_\Psi\right) \right] -1 } \right\}   
\nonumber \\  
&=& 2 g_\Phi \left( \frac{m_\Phi T}{2 \pi} \right)^{3/2}   
{\rm Li}_{3/2} \left( e^{\beta \bar{\mu}_\Phi} \right) +   
g \left( \frac{m_\Psi T}{2 \pi} \right)^{3/2}   
{\rm Li}_{3/2} \left( e^{\beta \bar{\mu}_\Psi} \right) \;,  
\label{SigmaPhi2}  
\end{eqnarray}  
and similarly for $\Sigma_{\Psi,\Psi}$,
  
\begin{eqnarray}  
\Sigma_{\Psi,\Psi} \equiv \Sigma_{\psi_1,\psi_1}=\Sigma_{\psi_2,\psi_2} &=&   
2 g_\Psi \left( \frac{m_\Psi T}{2 \pi} \right)^{3/2}   
{\rm Li}_{3/2} \left( e^{\beta \bar{\mu}_\Psi} \right) +   
g \left( \frac{m_\Phi T}{2 \pi} \right)^{3/2}   
{\rm Li}_{3/2} \left( e^{\beta \bar{\mu}_\Phi} \right) \;,  
\label{SigmaPsi2}  
\end{eqnarray}  
  
\noindent  
where ${\rm Li}_{3/2}(z)$ is a polylogarithmic function,
  
\begin{equation}  
{\rm Li}_{\alpha} (z) = \sum_{l=1}^{\infty} \frac{z^l}{l^{\alpha}}\;.  
\end{equation}  
  
The pressure in this case, $P(T,\mu_\Phi,\mu_\Psi)= - V_{\rm
  eff}(T,\phi_0=0,\psi_0=0)$, becomes just
  
\begin{eqnarray}  
P(T,\mu_\Phi,\mu_\Psi)&=&  \left( \frac{m_\Phi}{2 \pi} \right)^{3/2}  
T^{5/2} {\rm Li}_{5/2} \left( e^{\beta \bar{\mu}_\Phi} \right) +  
\left( \frac{m_\Psi}{2 \pi} \right)^{3/2}  
T^{5/2} {\rm Li}_{5/2} \left( e^{\beta \bar{\mu}_\Psi} \right)  
\nonumber \\  
&+& \frac{1}{2} \Sigma_{\Phi,\Phi} \left( \frac{m_\Phi T}{2 \pi} \right)^{3/2}   
{\rm Li}_{3/2} \left( e^{\beta \bar{\mu}_\Phi} \right)  
+ \frac{1}{2} \Sigma_{\Psi,\Psi} \left( \frac{m_\Psi T}{2 \pi} \right)^{3/2}   
{\rm Li}_{3/2} \left( e^{\beta \bar{\mu}_\Psi} \right)\;,  
\label{Pnormal}  
\end{eqnarray}  
  
\noindent  
where the two last terms in Eq. (\ref{Pnormal}) come from the sum of
two-bubble vacuum diagrams made of the quartic self-interaction and
cross-interaction vertices with the two additional terms, proportional to the
self-energies, added to the original Lagrangian and regarded as additional
interaction terms (the last term in Eq. (\ref{newL}) and the similar
contribution for the components of $\Psi$).  Using Eqs. (\ref{SigmaPhi2}) and
(\ref{SigmaPsi2}) in (\ref{Pnormal}), we obtain
  
\begin{eqnarray}  
P(T,\mu_\Phi,\mu_\Psi)&=&  \left( \frac{m_\Phi}{2 \pi} \right)^{3/2}  
T^{5/2} {\rm Li}_{5/2} \left( e^{\beta \bar{\mu}_\Phi} \right) +  
\left( \frac{m_\Psi}{2 \pi} \right)^{3/2}  
T^{5/2} {\rm Li}_{5/2} \left( e^{\beta \bar{\mu}_\Psi} \right)  
\nonumber \\  
&+& g_\Phi \left( \frac{m_\Phi T}{2 \pi} \right)^{3}   
{\rm Li}_{3/2}^2 \left( e^{\beta \bar{\mu}_\Phi} \right)  
+ g_\Psi \left( \frac{m_\Psi T}{2 \pi} \right)^{3}   
{\rm Li}_{3/2}^2 \left( e^{\beta \bar{\mu}_\Psi} \right)  
\nonumber \\  
&+& g (m_\Phi m_\Psi)^{3/2} \left(\frac{T}{2 \pi}\right)^3   
{\rm Li}_{3/2} \left( e^{\beta \bar{\mu}_\Phi} \right)  
{\rm Li}_{3/2} \left( e^{\beta \bar{\mu}_\Psi} \right)\;.  
\label{Pnormal2}  
\end{eqnarray}  
  
{}From Eq. (\ref{Pnormal2}) we can now fix the chemical potentials from the
expressions giving the densities,
  
\begin{eqnarray}  
\rho_\Phi &=& \frac{\partial P}{\partial \mu_\Phi} =  
\left( \frac{m_\Phi T}{2 \pi} \right)^{3/2}  
{\rm Li}_{3/2} \left( e^{\beta \bar{\mu}_\Phi} \right)  
\left[ (1 + 2 g_\Phi A_\Phi)   
\frac{\partial \bar{\mu}_\Phi}{\partial \mu_\Phi}  
+ g A_\Psi \frac{\partial \bar{\mu}_\Psi}{\partial \mu_\Phi}  
\right] \nonumber \\  
&+& \left( \frac{m_\Psi T}{2 \pi} \right)^{3/2}  
{\rm Li}_{3/2} \left( e^{\beta \bar{\mu}_\Psi} \right)  
\left[ (1 + 2 g_\Psi A_\Psi)   
\frac{\partial \bar{\mu}_\Psi}{\partial \mu_\Phi}  
+ g A_\Phi \frac{\partial \bar{\mu}_\Phi}{\partial \mu_\Phi}  
\right]\;,  
\label{rhoPhiN}  
\end{eqnarray}  
and
  
\begin{eqnarray}  
\rho_\Psi &=& \frac{\partial P}{\partial \mu_\Psi} =  
\left( \frac{m_\Psi T}{2 \pi} \right)^{3/2}  
{\rm Li}_{3/2} \left( e^{\beta \bar{\mu}_\Psi} \right)  
\left[ (1 + 2 g_\Psi A_\Psi)   
\frac{\partial \bar{\mu}_\Psi}{\partial \mu_\Psi}  
+ g A_\Phi \frac{\partial \bar{\mu}_\Phi}{\partial \mu_\Psi}  
\right] \nonumber \\  
&+& \left( \frac{m_\Phi T}{2 \pi} \right)^{3/2}  
{\rm Li}_{3/2} \left( e^{\beta \bar{\mu}_\Phi} \right)  
\left[ (1 + 2 g_\Phi A_\Phi)   
\frac{\partial \bar{\mu}_\Phi}{\partial \mu_\Psi}  
+ g A_\Psi \frac{\partial \bar{\mu}_\Psi}{\partial \mu_\Psi}  
\right]\;,  
\label{rhoPsiN}  
\end{eqnarray}  
where
  
\begin{equation}  
A_i = \frac{1}{T} \left( \frac{m_i T}{2 \pi} \right)^{3/2}  
{\rm Li}_{1/2} \left( e^{\beta \bar{\mu}_i} \right)\;.  
\label{Ai}  
\end{equation}  
  
\noindent  
The derivatives involving the chemical potentials in (\ref{rhoPhiN}) and
(\ref{rhoPsiN}), are defined by
  
\begin{eqnarray}  
&& \frac{\partial \bar{\mu}_\Phi}{\partial \mu_\Phi} =  
1-2 g_\Phi A_\Phi \frac{\partial \bar{\mu}_\Phi}{\partial \mu_\Phi}  
- g A_\Psi \frac{\partial \bar{\mu}_\Psi}{\partial \mu_\Phi}\;,  
\nonumber \\  
&& \frac{\partial \bar{\mu}_\Psi}{\partial \mu_\Phi} =  
- 2 g_\Psi A_\Psi \frac{\partial \bar{\mu}_\Psi}{\partial \mu_\Phi}  
- g A_\Phi \frac{\partial \bar{\mu}_\Phi}{\partial \mu_\Phi}\;,  
\nonumber \\  
&&  \frac{\partial \bar{\mu}_\Psi}{\partial \mu_\Psi} =  
1- 2 g_\Psi A_\Psi \frac{\partial \bar{\mu}_\Psi}{\partial \mu_\Psi}  
- g A_\Phi \frac{\partial \bar{\mu}_\Phi}{\partial \mu_\Psi}\;,  
\nonumber \\  
&& \frac{\partial \bar{\mu}_\Phi}{\partial \mu_\Psi} =  
- 2 g_\Phi A_\Phi \frac{\partial \bar{\mu}_\Phi}{\partial \mu_\Psi}  
- g A_\Psi \frac{\partial \bar{\mu}_\Psi}{\partial \mu_\Psi}\;.  
\label{set}  
\end{eqnarray}  
  
\noindent  
Eq. (\ref{set}) represents a set of equations for the derivatives of the
effective chemical potentials that can be easily solved and then the results
substituted back in Eqs. (\ref{rhoPhiN}) and (\ref{rhoPsiN}). The resulting
expressions are just
  
\begin{equation}  
\rho_\Phi = \left( \frac{m_\Phi T}{2 \pi} \right)^{3/2}  
{\rm Li}_{3/2} \left( e^{\beta \bar{\mu}_\Phi} \right)\;,  
\label{rhoPhiN2}  
\end{equation}  
and
  
\begin{equation}  
\rho_\Psi = \left( \frac{m_\Psi T}{2 \pi} \right)^{3/2}  
{\rm Li}_{3/2} \left( e^{\beta \bar{\mu}_\Psi} \right)\;.  
\label{rhoPsiN2}  
\end{equation}  
  
\noindent  
Eqs. (\ref{rhoPhiN2}) and (\ref{rhoPsiN2}) can also be used in the equations
(at the one-loop level) defining the self-energies in the normal phase, Eqs.
(\ref{SigmaPhi2}) and (\ref{SigmaPsi2}). Then, by expanding Eqs.
(\ref{rhoPhiN2}) and (\ref{rhoPsiN2}) in the high temperature limit
($\bar{\mu}_i/T \ll 1$) one obtains the expression for the chemical potentials
in terms of the densities (after inverting Eqs. (\ref{rhoPhiN2}) and
(\ref{rhoPsiN2})),
  
\begin{equation}  
\mu_\Phi \simeq 2 g_\Phi \rho_\Phi + g \rho_\Psi - \frac{T}{4 \pi}  
\left[ \left(\frac{2 \pi}{m_\Phi T}\right)^{3/2} \rho_\Phi -   
\zeta(3/2) \right]^2\;,  
\label{muPhiN}  
\end{equation}  
and
  
\begin{equation}  
\mu_\Psi \simeq 2 g_\Psi \rho_\Psi + g \rho_\Phi - \frac{T}{4 \pi}  
\left[ \left(\frac{2 \pi}{m_\Psi T}\right)^{3/2} \rho_\Psi -   
\zeta(3/2) \right]^2\;.  
\label{muPsiN}  
\end{equation}  
  
\noindent  
The above results, for the uncoupled two Bose gas ($g=0$) can be shown to
agree with the results of Ref. \cite{norway} obtained for the one Bose gas
case and similar analysis that also follow form our results (\ref{muPhiN}) and
(\ref{muPsiN}).

\section{Conclusions}

We have considered a nonrelativistic model suitable to describe a system of
homogeneous dilute Bose gas composed by two different types of atoms. A survey
of the literature \cite{coupled} shows that there is a growing interest in
this type of systems.  In general, coupled systems show a richer phase
structure in comparison with uncoupled ones due to the presence of a cross
coupling \cite{nossoprd,prdjulia,jpa1,jpa2}.  {}For example, coupled
nonrelativistic systems under the influence of external fields, represented by
one body terms, may exhibit re-entrant phases \cite{prdjulia,jpa1}. The
appearance of such phenomena depends on the sign of the inter-species
coupling, being completely ruled out in single species models. Apart from this
fact, one also observes that the uncoupled and coupled models produce
different values for critical quantities such as $T_c$.
  
In this work, our main motivation was to check how the presence of a inter
species coupling would affect the qualitative and quantitative behavior of the
transition regarding homogeneous coupled Bose gases. With this purpose we have
evaluated the effective potential at finite temperature in a nonperturbative
fashion to one-loop. Due to the complexity of zero temperature contributions,
the complete evaluation was only possible in the approximation where both
atomic masses are approximately the same.  Our results show a dramatic
difference concerning the case studied in Ref. \cite {prdjulia} where the one
body term represents, e.g., external fields and the Bose gas case, considered
here, where these terms represent chemical potentials. Our present results
exclude the possibility of exotic transition patterns such as inverse symmetry
breaking and re-entrant phases arising in models relevant for BEC. This is
rather satisfactory since, intuitively, one expects that the BEC transition
for coupled gases should also observe the same simple pattern observed by
monoatomic gases. That is, the system smoothly goes from an unsymmetric phase
to a symmetric phase as soon as the critical temperature is reached. This is
so because the value of cross coupling always appears squared eliminating the
possible occurrence of ISB.  Then, our second concern was to check the
numerical values for the critical temperatures at which symmetry restoration
occurs. Numerically, we found that the $T_c$ values are insensitive to the
coupling and coincide with the monoatomic results (which at this level of
approximation corresponds to the standard ideal gas result). This can be
understood in the view of the Hugenholtz-Pines theorem, in which case our
one-loop nonperturbative approximations cannot probe the effects of
interactions at $T=T_c$. Nevertheless, in the region $0 \le T < T_c$ (where
our approximation is more reliable) the curve describing the transition is
seen to be influenced by the cross coupling.  Taking the two species as having
different densities, we have started at $T=0$ increasing the temperature and
observing a first transition in which the more dilute specie always reaches
the non-condensed phase. At this first critical point the system decouples,
the denser specie remains in the condensed phase while the less dense acts as
a thermal cloud.  Then, a second transition occurs with the denser specie
reaching the gas phase exactly at the $T_c$ observed in the corresponding
monoatomic case.  If the two species have the same density the unbroken phase
is reached at once, which is not surprising, remembering that we took the
atomic masses as being approximately the same. In summary, as in the
monoatomic case and due to the Hugenholtz-Pines theorem, only the shapes of
the curves describing the thermal behavior of the effective potential minima
are sensitive to the numerical values of the couplings. The cross coupling, in
particular, has a non negligible effect in this parameterization hinting in
the possibility that a shift will appear if one computes, at higher orders,
the critical temperatures for interacting uncoupled and coupled Bose gases.
  
Though in this work we have considered a rather idealized case, in comparison
to real experiments performed in BEC, by only working with a homogeneous model
and neglecting any non-homogeneity effects, we hope that the approach
developed here may still be useful in the analysis of the thermodynamics in
realistic Bose-Einstein condensation experiments with coupled atomic gases.
The results and field theory methods used here could be considered, for
instance, to be applicable to trapped atomic gases in the central region of
wide traps.
  
{}Finally we remark that the coupled field model studied here, Eq.
(\ref{NRL}), could actually be viewed as the nonrelativistic limit of a two
complex scalar fields $\Phi$ and $\Psi$, both with conserved charges, with
chemical potentials $\mu_{r,i}$, $i=\Phi,\Psi$.  In the nonrelativistic limit,
the chemical potentials appearing in Eq. (\ref{NRL}) should then be identified
with $\mu_i \equiv \mu_{r,i} - m_i$, which is the correct identification of a
nonrelativistic chemical potential \cite{kapusta}.  Some of the details of the
nonrelativistic limit of the corresponding relativistic action of a two-scalar
field model were already given in Refs.  \cite{prdjulia,jpa1}, while for the
one field model this was previously discussed in \cite{benson}. The
calculations and analysis performed in the present work could then be easily
extended to the relativistic problem and the combined effects of temperature
and finite densities for the corresponding phase diagram of the model be
studied. This would be of particular relevance for studies related to the
early universe phase transition problems as well as the current heavy-ion
collision experiments. We will report on the extention of the results
presented in this work to the relativistic model in a forthcoming publication.

\acknowledgments
  
The authors have been partially supported by CNPq-Brazil.  ROR would also like
to thank partial support from FAPERJ.

\end{document}